\documentclass[12pt]{article}
\usepackage{amsfonts}
\usepackage{amssymb}
\usepackage{amsmath}
\usepackage{amsthm}
\usepackage{color}
\usepackage{graphics}
\usepackage{epsfig,amssymb,latexsym,verbatim}
\usepackage{epstopdf}
\usepackage{multirow}
\usepackage{multicol}
\usepackage{float}
\usepackage{hyperref}
\usepackage{graphicx, amssymb}
\usepackage{natbib}
\usepackage{breakcites}

\floatstyle{ruled}
\newfloat{algorithm}{tbp}{loa}
\floatname{algorithm}{Algorithm}
\renewcommand{\baselinestretch} {1.2}

\newtheorem{theorem}{\textbf{Theorem}}
\newtheorem{corollary}{\textbf{Corollary}}
\newtheorem{proposition}{\textbf{Proposition}}

\newcommand{\blind}{0}

\begin{document}
\def\spacingset#1{\renewcommand{\baselinestretch}%
{#1}\small\normalsize} \spacingset{1}
\date{}

\thispagestyle{empty} \baselineskip=12pt

\if0\blind
{
\title{\textbf{A concave pairwise fusion approach to subgroup analysis}}

\author{Shujie Ma\thanks{The research of Ma is supported in part by the
U.S. NSF grant DMS-13-06972.}\\
Department of Statistics, University of California Riverside\\
and\\
Jian Huang\thanks{Corresponding author. The research of Huang is supported in part by the U.S. NSF grant DMS-12-08225.}\\
Department of Statistics and Actuarial Science, University of Iowa}
\maketitle
}\fi


\begin{center}
{\Large \textbf{Abstract}}
\end{center}

An important step in developing individualized treatment strategies is to correctly identify subgroups of a heterogeneous population, so that specific treatment can be given to each subgroup. In this paper, we consider the situation with samples drawn from a population consisting of subgroups with different means, along with certain covariates. We propose a penalized approach for subgroup analysis based on a regression model, in which heterogeneity is driven by unobserved latent factors and thus can be represented by using subject-specific intercepts. We apply concave penalty functions to pairwise differences of the intercepts. This procedure automatically divides the observations into subgroups.
We develop an alternating direction method of multipliers algorithm with concave penalties to implement the proposed approach and demonstrate its convergence. We also establish the theoretical properties of our proposed estimator and determine the order requirement of the minimal difference of signals between groups in order to recover them. These results provide a sound basis for making statistical inference in subgroup analysis. Our proposed method is further illustrated by simulation studies and analysis of the Cleveland heart disease dataset.

\baselineskip=12pt


\bigskip\bigskip\noindent%

\bigskip\noindent
\textit{Short title:} Subgroup analysis

\vfill


\newpage
\spacingset{1.45} 

\section{Introduction\label{SEC:Intro}}

Personalized medicine has gained much attention in the past decade, which
emphasizes the use of information available on individual patients to make
treatment decisions. Developing individualized treatment strategies requires
sophisticated analytic tools. 
One of the key statistical challenges is to correctly identify subgroups from a
heterogeneous population, so that specific medical therapies can be given to
each subgroup. A popular method for analyzing data from a
heterogeneous population is to view data as coming from a mixture of
subgroups with their own sets of parameter values
and then use finite mixture model analysis \citep{Everitt.Hand:1981}. The
mixture model approach has been widely used for data clustering and
classification; see \cite{Banfield.Raftery:1993}, \cite%
{Hastie.Tibshirani:1996}, \cite{McNicholas:2010} and \cite{wei.kosorok:2013}
for the Gaussian mixture model approaches, \cite{shen.he:2015} for a
logistic-normal mixture model method, and \cite{Chaganty.Liang:2013} for a
low-rank method for mixtures of linear regressions which provides a good initialization for the EM algorithm typically used in estimation of mixture models. The
mixture model-based approach as a supervised clustering method needs to
specify an underlying distribution for the data, and it also requires specifying the
number of mixture components in the population which is often difficult to do in
practice.

In this paper, we propose a new approach to automatically detecting and
identifying homogeneous subgroups based on a concave pairwise fusion penalty
without the knowledge of an \textit{a priori} classification or a natural
basis of separating a sample into subsets. Let $y_{i}$ be the response
variable for the $i^{\text{th}}$ subject. After adjusting for the effects of
a set of covariates $\mathbf{x}_{i}=(x_{i1},\ldots ,x_{ip})^{\text{T}}$, we
consider subgroup analysis for $\mathbf{y=(}y_{1},\ldots ,y_{n})^{\text{T}}$
with the heterogeneity driven by unknown or unobserved latent factors, which
can be modeled through subject-specific intercepts in regression. Hence, we
consider
\begin{equation}
y_{i}=\mu _{i}+\mathbf{x}_{i}^{\text{T}}\mathbf{\boldsymbol{\beta }}\mathbf{+%
}\epsilon _{i},i=1,\ldots ,n,  \label{Mod1}
\end{equation}%
where $\mu _{i}$'s are unknown subject-specific intercepts, $\mathbf{%
\boldsymbol{\beta }}\mathbf{=(}\beta _{1},\ldots ,\beta _{p})^{\text{T}}$ is
the vector of unknown coefficients for the covariates $\mathbf{x}_{i}$, and $%
\epsilon _{i}$ is the error term independent of $\mathbf{x}_{i}$ with $%
E(\epsilon _{i})=0$ and Var$(\epsilon _{i})=\sigma ^{2}$. For example, in
biomedical studies, $y_{i}$ can be certain phenotype associated with some
disease such as the maximal heart rate which is related to cardiac mortality
or body mass index associated with obesity, and $\mathbf{x}_{i}$ is a set of
observed covariates such as gender, age, race, etc. After adjusting for the
effects of the covariates, the distribution of the response is still
heterogeneous, as demonstrated by multiple modes in the density plot shown in
Figure \ref{FIG:density} for our heart disease application. This
heterogeneity can be caused by unobserved latent factors, so that it is
modeled through the subject-specific $\mu _{i}$'s.

It is worth noting that if the factors contributing to this heterogeneity, for example, different treatments, become
available, then $\mu
_{i}$ can be written as $\mu _{i}=\mu +\mathbf{z}_{i}^{\text{T}}\mathbf{%
\boldsymbol{\theta }}$, where $\mathbf{z}_{i}$ are the observed variables
for the treatments and $\mathbf{\boldsymbol{\theta }}$ are the coefficients of $\mathbf{z}_{i}$. One interesting application in personalized medicine
is that the coefficients for $\mathbf{z}_{i}$ can be subject-specific, since
the same treatment may have different effects on patients. For this case, we
can consider the model with heterogeneous effects of some covariates given
as
\begin{equation}
y_{i}=\mu +\mathbf{z}_{i}^{\text{T}}\mathbf{\boldsymbol{\theta }}_{i}+
\mathbf{x}_{i}^{\text{T}}\mathbf{\boldsymbol{\beta }}\mathbf{+}\epsilon
_{i},i=1,\ldots ,n.  \label{Mod2}
\end{equation}

Throughout this paper, we focus on studying model (\ref{Mod1}) by considering that
the heterogeneity comes from unobserved latent factors. However, our
proposed estimation method and the associated theoretical properties for
model (\ref{Mod1}) can be extended to model (\ref{Mod2}) with some
modifications. We provide the detailed estimation procedure for model
(\ref{Mod2}) in Section A.4 of the Supplemental Materials for interested readers. Assumptions of the structure are needed in order to estimate model (\ref{Mod1}). To this end, we
assume that $\mathbf{y=(}y_{1},\ldots ,y_{n})^{\text{T}}$ are from $K$
different groups with $K\geq 1$
and the data from the same group have the same intercept. In other words, let $
\mathcal{G=(G}_{1},\ldots ,\mathcal{G}_{K})$ be a partition of $\{1,\ldots
,n\}$. We have $\mu _{i}=\alpha _{k}$ for all $i\in \mathcal{G}_{k}$, where $
\alpha _{k}$ is the common value for the $\mu _{i}$'s from group $\mathcal{G}
_{k}$.
In practice, the number of groups $K$ is unknown. However, it is usually reasonable to
assume that $K$ is much smaller than $n$. Our goal is to
estimate $K$ and identify the subgroups. We are also interested in
estimating the intercepts $(\alpha _{1},\ldots ,\alpha _{K})$ and the
regression parameter $\boldsymbol{\beta }$. We propose a concave pairwise
fusion penalized least squares approach for this purpose and derive an
alternating direction method of multipliers (ADMM, \cite{Boyd:2011})
algorithm for implementing the proposed approach.

Several authors have studied the problem of exploring homogeneity effects
of covariates in the regression setting by assuming that the true
coefficients are divided into a few clusters with common values. For
instance, \cite{Tibshirani.Saunders:2005} proposed the fused LASSO method
which applies $L_{1}$ penalties to the pairs of adjacent coordinates given
that a complete ordering of covariates is available.
\cite{Bondell.Reich:2008} proposed the OSCAR method where a special octagonal
shrinkage penalty is applied to each pair of coordinates. \cite{Shen.Huang:2010}
developed a group pursuit approach with truncated $L_{1}$
penalties to the pairwise differences, and \cite{ke.fan.wu:2013} proposed a
method called CARDS. All the above methods are about estimating homogeneity
effects of covariates, which is different from our work aiming to identify
subgroups of the observations. \cite{Guo-etal:2010} proposed using a
pairwise $L_1$ fusion penalty for identifying informative variable in the
context of Gaussian model-based cluster analysis. In the unsupervised
learning setting, a recent paper (\cite{chi.lange:2014}) considered the
convex clustering problem 
and investigated the ADMM and the alternating minimization algorithms with
the convex $L_{p}$ ($p\geq 1$) penalties applied to the pairwise differences
of the data points.

The ADMM has good convergence properties for convex loss functions
with the $L_{p}$, $p\geq 1$,
penalties (\cite{Boyd:2011} and \cite{chi.lange:2014}).
Moreover, the $L_{1}$ penalty can shrink some pairwise differences of the parameter estimates to zero.
However, the $L_{1}$ penalty generates large biases of the estimates in
each iteration of the algorithm. As a result, it may not be able to
identify the subgroups, as illustrated in Figure \ref{FIG:path}. To address this issue, \cite{chi.lange:2014} propose to multiply
nonnegative weights to the $L_{1}$ norms to reduce the bias. However, the choice of the weights
can dramatically affect the quality of the clustering solution, and there is
no clear rule for how to choose the weights. Thus, a penalty which can produce
unbiased estimates is more desirable for identifying subgroups. We propose an ADMM algorithm by using concave pairwise fusion penalties for estimation of model (\ref{Mod1}). The concave penalties in the
optimization problem such as the smoothly clipped absolute deviations
penalty (SCAD, \cite{fan.li:2001}) and the minimax concave penalty (MCP,
\cite{zhang:2010}) enjoy the unbiasedness property. We then derive
the convergence properties of the ADMM algorithm. Moreover, we
provide theoretical analysis of the proposed estimators. Specifically, we
derive the order requirement of the minimum difference of signals between
groups in order to identify the true subgroups. We also establish the oracle property that under mild regularity conditions the oracle estimator is a local minimizer of the objective function with a high probability. The oracle estimator is obtained from least
squares regression by assuming that the true group structure is known.

The rest of this paper is organized as follows. In Section \ref{SEC:subgroup}
we describe the proposed approach in detail. In Section \ref{SEC:computation}
we derive an ADMM algorithm with concave penalties. We then state the
theoretical properties of the proposed approach in Section \ref{SEC:theory}.
In Sections \ref{SEC:examples} we evaluate the finite sample properties of
the proposed procedures via simulation studies. Section \ref%
{SEC:applications} illustrates the proposed method through a data example.
Some concluding remarks are given in Section \ref{SEC:Discussion}. The estimation procedure for model (\ref{Mod2}) and all the
technical proofs are provided in the on-line Supplemental Materials.

\section{Subgroup analysis via concave pairwise fusion}

\label{SEC:subgroup}

For estimation of model (\ref{Mod1}), we propose a concave pairwise
fusion penalized least squares approach. The objective function is
\begin{equation}
Q_{n}(\mathbf{\boldsymbol{\mu }}{,\mathbf{\boldsymbol{\beta }};\lambda })=%
\frac{1}{2}\sum\nolimits_{i=1}^{n}(y_{i}-\mu _{i}-\mathbf{x}_{i}^{\text{T}}%
\mathbf{\boldsymbol{\beta })}^{2}+\sum\nolimits_{1\leq i<j\leq n}p(|\mu
_{i}-\mu _{j}|,\lambda ),  \label{EQ:objective}
\end{equation}%
where $\mathbf{\boldsymbol{\mu }}=(\mu _{1},\ldots ,\mu _{n})^{\text{T}}$,
and $p(\cdot ,\lambda )$ is a concave penalty function with a tuning
parameter $\lambda \geq 0$.

For a given $\lambda >0$, define
\begin{equation*}
(\widehat{\mathbf{\boldsymbol{\mu }}}(\lambda ),\widehat{\mathbf{\boldsymbol{%
\beta }}}(\lambda ))=\hbox{argmin}_{\mu, \boldsymbol{\beta}} \,Q_{n}(\mathbf{%
\boldsymbol{\mu }}{,\mathbf{\boldsymbol{\beta }};\lambda }).
\end{equation*}%
The penalty shrinks some of the pairs $\mu _{j}-\mu _{k}$ to zero. Based on
this, we can partition the sample into subgroups. Specifically, let $%
\widehat{{\lambda }}$ be the value of the tuning parameter selected based on
a data-driven procedure such as the BIC. For simplicity, write $(\widehat{%
\mathbf{\boldsymbol{\mu }}},\widehat{\mathbf{\boldsymbol{\beta }}})\equiv (%
\widehat{\mathbf{\boldsymbol{\mu }}}(\widehat{{\lambda }}),\widehat{\mathbf{%
\boldsymbol{\beta }}}(\widehat{{\lambda }}))$. Let $\{\widehat{\alpha }%
_{1},\ldots ,\widehat{\alpha }_{\widehat{K}}\}$ be the distinct values of $%
\widehat{\mathbf{\boldsymbol{\mu }}}$. Let $\widehat{\mathcal{G}}_{k}=\{i:%
\widehat{\mu }_{i}=\widehat{\alpha }_{k},1\leq i\leq n\},1\leq k\leq
\widehat{K}$. Then $\{\widehat{\mathcal{G}}_{1},\ldots ,\widehat{\mathcal{G}}%
_{\widehat{K}}\}$ constitutes a partition of $\{1,\ldots ,n\}$.

An important question is which penalty function should be used here. The $%
L_{1}$ penalty with $p_{\gamma }(t,\lambda )=\lambda t$ applies the same
thresholding to all pairs $|\mu _{i}-\mu _{j}|$. As a result, it leads to
biased estimates and may not be able to correctly recover the subgroups.
This is similar to the situation in variable selection where the lasso tends
to over-shrink large coefficients. In our numerical
studies, we found that the $L_{1}$ penalty tends to either yield a large
number of subgroups or no subgroup on the solution path. %
%
Hence, a penalty which can produce unbiased estimates is more appealing.
This motivates us to use the concave penalties including the smoothly
clipped absolute deviation penalty (SCAD, \cite{fan.li:2001}) and the
minimax concave penalty (MCP, \cite{zhang:2010}). These penalties are
asymptotically unbiased and are more aggressive in enforcing a sparser
solution. Thus, they are better suited for the current problem, since the
number of subgroups is usually much smaller than the sample size.

The MCP has the form
\begin{equation*}
p_{\gamma}(t,\lambda )=\lambda \int_{0}^{t}(1-x/(\gamma \lambda ))_{+}dx,
\gamma > 1,
\end{equation*}
and the SCAD penalty is
\begin{equation*}
p_{\gamma}(t,\lambda )=\lambda \int_{0}^{t}\min \{1,(\gamma -x/\lambda
)_{+}/(\gamma -1)\}dx, \gamma > 2,
\end{equation*}
where $\gamma$ is a parameter that controls the concavity of the penalty
functions. In particular, both penalties converge to the $L_1$ penalty as $%
\gamma \to \infty$. Here and in the rest of the paper, we put $\gamma$ in the subscript
to indicate the dependence of these penalty functions on it. Following \cite%
{fan.li:2001} and \cite{zhang:2010}, we treat $\gamma$ as a fixed constant.
These concave penalties enjoy the sparsity as the $L_{1}$ penalty that it
can automatically yield zero estimates. More importantly, it has the
unbiasedness property in that it does not shrink large estimated parameters, so
that they remain unbiased in the iterations. This property is particularly
essential in the ADMM algorithms since the biases in the iterations may
significantly affect the search for subgroups.

\section{Computation}


\label{SEC:computation}

It is difficult to compute the estimates directly by minimizing the
objective function (\ref{EQ:objective}) due to the fact that the penalty
function is not separable in $\mu _{i}$'s. We reparameterize the criterion
by introducing a new set of parameters $\eta _{ij}=\mu _{i}-\mu _{j}$. Then
the minimization of (\ref{EQ:objective}) is equivalent to the constraint
optimization problem,
\begin{gather}
S(\mathbf{\boldsymbol{\mu} },\mathbf{\boldsymbol{\beta} ,\boldsymbol{\eta} )=%
}\frac{1}{2}\sum\nolimits_{i=1}^{n}(y_{i}-\mu _{i}-\mathbf{x}_{i}^{\text{T}}%
\mathbf{\boldsymbol{\beta} )}^{2}+\sum\nolimits_{i<j}p_{\gamma }(|\eta
_{ij}|,\lambda ),  \notag \\
\text{subject to }\mu _{i}-\mu _{j}-\eta _{ij}=0,  \label{EQ:opt}
\end{gather}%
where $\mathbf{\boldsymbol{\eta} }=\{\eta _{ij},i<j\}$. By the augmented
Lagrangian method (ALM), the estimates of the parameters can be obtained by
minimizing
\begin{equation}
L(\mathbf{\boldsymbol{\mu} },\mathbf{\boldsymbol{\beta} ,\boldsymbol{\eta} },%
\mathbf{\boldsymbol{\upsilon} })=S(\mathbf{\boldsymbol{\mu} },\mathbf{%
\boldsymbol{\beta} ,\boldsymbol{\eta} })+\sum\nolimits_{i<j}\upsilon
_{ij}(\mu _{i}-\mu _{j}-\eta _{ij})+\frac{\vartheta }{2}\sum\nolimits_{i<j}(%
\mu _{i}-\mu _{j}-\eta _{ij}\mathbf{)}^{2},  \label{EQ:L}
\end{equation}%
where the dual variables $\boldsymbol{\upsilon} =\{\upsilon _{ij},i<j\}$ are
Lagrange multipliers and $\vartheta$ is the penalty parameter. We compute
the estimators of $(\boldsymbol{\mu}, \boldsymbol{\beta},\boldsymbol{\eta} ,%
\boldsymbol{\upsilon} )$ through iterations by the ADMM.

It is noteworthy that by using the concave penalties, although the objective
function $L(\mathbf{\boldsymbol{\mu} },\mathbf{\boldsymbol{\beta} ,%
\boldsymbol{\eta} ,\boldsymbol{\upsilon} )}$ is not a convex function, it is
convex with respect to each $\eta _{ij}$ when $\gamma >1/\vartheta $ for the
MCP penalty and $\gamma >1/\vartheta +1$ for the SCAD penalty. Moreover, for
given $(\mathbf{\boldsymbol{\mu} },\mathbf{\boldsymbol{\beta} ,\boldsymbol{
\eta} ,\boldsymbol{\upsilon} )} $, the minimizer of $L(\mathbf{\boldsymbol{
\mu} },\mathbf{\boldsymbol{\beta} ,\boldsymbol{\eta} ,\boldsymbol{\upsilon} )
}$ with respect to $\eta _{ij}$ is unique and has a closed-form expression
for the $L_{1}$, MCP and SCAD penalties, respectively. Specifically, for
given $(\mathbf{\boldsymbol{\mu} },\mathbf{\boldsymbol{\beta} ,\boldsymbol{
\eta} ,\boldsymbol{\upsilon} )}$, the minimization problem is the same as
minimizing
\begin{equation}
\frac{\vartheta }{2}(\delta _{ij}-\eta _{ij}\mathbf{)}^{2}+p_{\gamma }(|\eta
_{ij}|,\lambda )  \label{EQ:etaij}
\end{equation}%
with respect to $\eta _{ij}$, where $\delta _{ij}=\mu _{i}-\mu
_{j}+\vartheta ^{-1}\upsilon _{ij}$. Hence, the closed-form solution for the
$L_{1}$ penalty is
\begin{equation}
\widehat{\eta }_{ij}=\text{ST}(\delta _{ij},\lambda/\vartheta ),
\label{EQ:etalasso}
\end{equation}%
where ST$(t,\lambda )=$sign$(t)(\left\vert t\right\vert -\lambda )_{+}$ is
the soft thresholding rule, and $(x)_{+}=x$ if $x>0$, and $(x)_{+}=0$ otherwise. For
the MCP penalty with $\gamma >1/\vartheta $, it is
\begin{equation}
\widehat{\eta }_{ij}=\left\{
\begin{array}{cc}
\frac{\text{ST}(\delta _{ij},\lambda /\vartheta )}{1-1/(\gamma \vartheta )}
& \text{ if }|\delta _{ij}|\leq \gamma \lambda \\
\delta _{ij} & \text{ if }|\delta _{ij}|>\gamma \lambda%
\end{array}%
\right. .  \label{EQ:etaMCP}
\end{equation}%
For the SCAD penalty with $\gamma >1/\vartheta +1$, it is
\begin{equation}
\widehat{\eta }_{ij}=\left\{
\begin{array}{cc}
\text{ST}(\delta _{ij},\lambda /\vartheta ) & \text{ if }|\delta _{ij}|\leq
\lambda +\lambda /\vartheta \\
\frac{\text{ST}(\delta _{ij},\gamma \lambda /((\gamma -1)\vartheta ))}{%
1-1/((\gamma -1)\vartheta )} & \text{ if }\lambda +\lambda /\vartheta <\text{
}|\delta _{ij}|\leq \gamma \lambda \\
\delta _{ij} & \text{if }|\delta _{ij}|>\gamma \lambda%
\end{array}%
\right. .  \label{EQ:etaSCAD}
\end{equation}

\subsection{Algorithm\label{SEC:algorithm}}

We now describe the computational algorithm based on the ADMM for minimizing
the objective function (\ref{EQ:objective}). It consists of steps for iteratively
updating $\mathbf{\boldsymbol{\ \mu }},\mathbf{\boldsymbol{\beta },%
\boldsymbol{\eta }}$ and $\mathbf{\boldsymbol{\upsilon }}$. Denote the $L_{2}
$ norm of any vector $\mathbf{a}$ by $||\mathbf{a||}$. The main ingredients
of the algorithm are as follows.

First, for a given $(\mathbf{\boldsymbol{\eta} ,\boldsymbol{\upsilon} )}$,
to obtain an update of $\mathbf{\boldsymbol{\mu} }$ and $\mathbf{\boldsymbol{%
\beta} }$, we set the derivatives \newline ${\partial L(\mathbf{\boldsymbol{\mu} },%
\mathbf{\boldsymbol{\beta} ,\boldsymbol{\eta} ,\boldsymbol{\upsilon} )}}/{%
\partial \mathbf{\boldsymbol{\mu} }}$ and ${\partial L(\mathbf{\boldsymbol{%
\mu} },\mathbf{\boldsymbol{\beta} ,\boldsymbol{\eta} ,\boldsymbol{\upsilon} )%
}}/{\partial \mathbf{\boldsymbol{\beta} }}$ to zero, where
\begin{eqnarray}
L(\mathbf{\boldsymbol{\mu} },\mathbf{\boldsymbol{\beta} ,\boldsymbol{\eta} ,%
\boldsymbol{\upsilon} )} &=&\frac{1}{2}\sum\nolimits_{i=1}^{n}(y_{i}-\mu
_{i}-\mathbf{x}_{i}^{\text{T}}\mathbf{\boldsymbol{\beta} )}^{2}+\frac{%
\vartheta }{2}\sum\nolimits_{i<j}\{(e_{i}-e_{j})^{\text{T}}\mathbf{%
\boldsymbol{\mu} -}\eta _{ij}+\vartheta ^{-1}\upsilon _{ij}\}^{2}+C  \notag
\\
&=&\frac{1}{2}\left\Vert \mathbf{\boldsymbol{\mu} -y+X\boldsymbol{\beta} }%
\right\Vert ^{2}+\frac{\vartheta }{2}\left\Vert \mathbf{\boldsymbol{\Delta}
\boldsymbol{\mu} -\mathbf{\boldsymbol{\eta} +}}\vartheta ^{-1}\mathbf{%
\boldsymbol{\upsilon} }\right\Vert ^{2}+C.  \label{EQ:Lmubeta}
\end{eqnarray}%
Here $C$ is a constant independent of $\mathbf{\boldsymbol{\mu} }$ and $%
\mathbf{\boldsymbol{\beta} }$, $\mathbf{y=(}y_{1},\ldots ,y_{n})^{\text{T}}$%
, $\mathbf{X=(x}_{1},\ldots ,\mathbf{x}_{n})^{\text{T}}$, $e_{i}$ is the $i$%
th unit $n\times 1$ vector whose $i$th element is 1 and the remaining ones
are 0,
and $\boldsymbol{\Delta}\mathbf{=\{}(e_{i}-e_{j}),i<j\}^{\text{T}}$. Thus,
for given $\mathbf{\mathbf{\boldsymbol{\eta} }}^{(m)}$ and $\mathbf{%
\boldsymbol{\upsilon} }^{(m)}$ at the $m^{ \text{th}}$ step, the updates $%
\mathbf{\boldsymbol{\mu} }^{(m+1)}$ and $\mathbf{\boldsymbol{\beta} }
^{(m+1)}$, which are the minimizers of $L(\mathbf{\boldsymbol{\mu} },\mathbf{%
\boldsymbol{\beta} , \mathbf{\boldsymbol{\eta} }}^{(m)},\mathbf{\boldsymbol{%
\upsilon} }^{(m)})$, are
\begin{equation*}
\mathbf{\boldsymbol{\mu}}^{(m+1)}\mathbf{=(I+}\vartheta \mathbf{\boldsymbol{%
\Delta} }^{\text{T}} \mathbf{\boldsymbol{\Delta} -Q}_{\mathbf{X}}\mathbf{)}%
^{-1}\{\mathbf{(I-Q}_{\mathbf{X}} \mathbf{)y+}\vartheta \mathbf{\boldsymbol{%
\Delta} }^{\text{T}}(\mathbf{\mathbf{\boldsymbol{\eta} }} ^{(m)}\mathbf{-}%
\vartheta ^{-1}\mathbf{\boldsymbol{\upsilon} }^{(m)}\mathbf{)\},}
\end{equation*}
where $\mathbf{Q}_{\mathbf{X}}=\mathbf{X(X}^{\text{T}}\mathbf{X)}^{-1}%
\mathbf{X}^{\text{T}}$, and
\begin{equation*}
\mathbf{\boldsymbol{\beta} }^{(m+1)}=\mathbf{(X}^{\text{T}}\mathbf{X)}^{-1}%
\mathbf{X}^{\text{T}}(\mathbf{y-\boldsymbol{\mu} }^{(m+1)}).
\end{equation*}%
We further can derive $\mathbf{I+}\vartheta \boldsymbol{\Delta} ^{\text{T}}%
\boldsymbol{\Delta} \mathbf{=}(1+n\vartheta )\mathbf{I-}\vartheta \mathbf{11}%
^{\text{T}}$.

Second, the update of $\eta _{ij}$ at the $(m+1)^{\text{th}}$ iteration is
obtained by the formula given in (\ref{EQ:etalasso}), (\ref{EQ:etaMCP}) and (%
\ref{EQ:etaSCAD}), respectively, by the Lasso, MCP and SCAD penalties with $%
\delta _{ij}$ replaced by $\delta _{ij}^{(m+1)}=\mu _{i}^{(m+1)}-\mu
_{j}^{(m+1)}+\vartheta ^{-1}\upsilon _{ij}^{(m)}$.

Finally, the estimate of $\upsilon _{ij}$ is updated as
\begin{equation*}
\upsilon _{ij}^{(m+1)}=\upsilon _{ij}^{(m)}+\vartheta (\mu _{i}^{(m+1)}-\mu
_{j}^{(m+1)}-\eta _{ij}^{(m+1)}\mathbf{).}
\end{equation*}

Based on the above discussion, the algorithm consists of the following steps:

Step 1. Find initial estimates $\mathbf{\boldsymbol{\beta }}^{(0)}$ from
least squares regression\ by letting $\mu _{i}=\mu $ for all $i$. Let the
initial estimates $\mathbf{\boldsymbol{\mu }}^{(0)}=\mathbf{y-X\boldsymbol{%
\beta }}^{(0)}$, $\eta _{ij}^{(0)}=\mu _{i}^{(0)}-\mu _{j}^{(0)}$ and $%
\mathbf{\upsilon }^{(0)}=\mathbf{0}$.

Step 2. At iteration $m+1$, compute ($\mathbf{\boldsymbol{\mu} }^{(m+1)},%
\mathbf{\boldsymbol{\beta} }^{(m+1)},\mathbf{\mathbf{\boldsymbol{\eta} }}%
^{(m+1)},\mathbf{\boldsymbol{\upsilon} }^{(m+1)})$ by the methods described
above.

Step 3. Terminate the algorithm if the stopping rule is met at step $m+1$.
Then ($\mathbf{\boldsymbol{\mu} }^{(m+1)},\mathbf{\boldsymbol{\beta} }%
^{(m+1)}$, $\mathbf{\boldsymbol{\eta} }^{(m+1)},\mathbf{\boldsymbol{\upsilon}
}^{(m+1)})$ are our final estimates ($\widehat{\mathbf{\boldsymbol{\mu} }},%
\widehat{\mathbf{\boldsymbol{\beta} }},\widehat{\mathbf{\boldsymbol{\eta} }},%
\widehat{\mathbf{\boldsymbol{\upsilon} }})$. Otherwise, we go to Step 2.


\textbf{Remark 1. } We track the progress of the ADMM based on the primal
residual $\mathbf{r}^{(m+1)}=\boldsymbol{\Delta} \mathbf{\boldsymbol{\mu} }%
^{(m+1)} \mathbf{-\mathbf{\boldsymbol{\eta} }}^{(m+1)}$ . We stop the
algorithm when $\mathbf{r}^{(m+1)}$ is close to zero such that $\left\Vert
\mathbf{r} ^{(m+1)}\right\Vert <\epsilon $ for some small value $\epsilon $.

\textbf{Remark 2. } This algorithm enables us to have $\widehat{\eta }%
_{ij}=0 $ for a large $\lambda$. We put $y_{i}$ and $y_{j}$ in the same
group if $\widehat{\eta }_{ij}=0$. As a result, we have $\widehat{K}\ $%
estimated groups $\widehat{\mathcal{G}}_{1},\ldots ,\widehat{\mathcal{G}}_{%
\widehat{K} } $ and let the estimated intercept for the $k^{\text{th}}$
group be $\widehat{\alpha }_{k}=|\widehat{\mathcal{G}}_{k}|^{-1}\sum%
\nolimits_{i\in \widehat{\mathcal{G}}_{k}}\widehat{\mu }_{i}$, where $|%
\widehat{\mathcal{G}} _{k}|$ is the cardinality of $\widehat{\mathcal{G}}%
_{k} $.


\subsection{Convergence of the algorithm}

We next consider the convergence properties of the ADMM algorithm.

\begin{proposition}
\label{PROP:primal} The primal residual $\mathbf{r}^{(m)}=\boldsymbol{\Delta}
\mathbf{\boldsymbol{\mu}}^{(m)}\mathbf{-\mathbf{\boldsymbol{\eta} }}^{(m)}$
and the dual residual $\mathbf{s}^{(m+1)}=\vartheta \mathbf{\boldsymbol{%
\Delta} }^{\text{T}}(\mathbf{\boldsymbol{\eta} }^{(m+1)}-\mathbf{\boldsymbol{%
\eta} }^{(m)})$ of the ADMM satisfy that $\lim_{m\rightarrow \infty }||%
\mathbf{r}^{(m)}||^{2}=0$ and $\lim_{m\rightarrow \infty }||\mathbf{s}%
^{(m)}||^{2}=0$ for both of the MCP and SCAD\ penalties.
\end{proposition}

The proof of this result is given in the online supplement.
Proposition \ref{PROP:primal} shows that the primal feasibility and dual
feasibility are achieved by the algorithm. Therefore, it converges to an
optimal point. This optimal point may be a local minimum of the objective
function when a concave penalty function is applied.

\section{Theoretical properties\label{SEC:theory}}

In this section, we study the theoretical properties of the proposed
estimator based on concave penalty functions. Specifically, we derive the
order requirement of the minimum difference of signals between groups in
order to recover the true groups and the oracle property that under some
regularity conditions the oracle estimator is a local minimizer of the objective function with a high
probability. Let $\mathcal{M}_{\mathcal{G}}$ be the subspace of $R^{n}$,
defined as
\begin{equation*}
\mathcal{M}_{\mathcal{G}}=\{\mathbf{\boldsymbol{\mu }\in }R^{n}:\mu _{i}=\mu
_{j}\text{, for any }i,j\in \mathcal{G}_{k},1\leq k\leq K\}.
\end{equation*}%
For each $\mathbf{\boldsymbol{\mu }\in }\mathcal{M}_{\mathcal{G}}$, it can
be written as $\mathbf{\boldsymbol{\mu }=Z\boldsymbol{\alpha }}$, where $%
\mathbf{Z=\{}z_{ik}\}$ is the $n\times K$ matrix with $z_{ik}=1$ for $i\in
\mathcal{G}_{k\text{ }}$and $z_{ik}=0$ otherwise, and ${\boldsymbol{\alpha }}
$ is a $K\times 1$ vector of parameters. By matrix calculation, we have
$\mathbf{D=Z}^{\text{T}}\mathbf{Z=}\text{diag}(\left\vert \mathcal{G}%
_{1}\right\vert ,\ldots ,\left\vert \mathcal{G}_{K}\right\vert ),$ where $%
\left\vert \mathcal{G}_{k}\right\vert $ denotes the number of elements in $%
\mathcal{G}_{k}$. Define $\left\vert \mathcal{G}_{\min }\right\vert \mathcal{%
=}\min_{1\leq k\leq K}\left\vert \mathcal{G}_{k}\right\vert $ and $%
\left\vert \mathcal{G}_{\max }\right\vert \mathcal{=}\max_{1\leq k\leq
K}\left\vert \mathcal{G}_{k}\right\vert $. Let $\mathbf{\ X=(X}_{1},\ldots
\mathbf{X}_{p})$, where $\mathbf{X}_{j}$ is the $j$th column of $\mathbf{X}$%
. Denote
\begin{equation*}
\rho (t)=\lambda ^{-1}p_{\gamma }(t,\lambda )\ \mbox{ and }\ \overline{\rho }%
(t)=\rho ^{\prime }(|t|)\hbox{sgn}(t).
\end{equation*}%
For any vector $\mathbf{\boldsymbol{\zeta }}=\left( \zeta _{1},\ldots ,\zeta
_{s}\right) ^{\text{T}}\in R^{s}$, denote $\left\Vert \mathbf{\boldsymbol{%
\zeta }}\right\Vert _{\infty }=\max_{1\leq l\leq s}\left\vert \zeta
_{l}\right\vert $. For any symmetric matrix $\mathbf{A}_{s\times s}$, denote
its $L_{2}$ norm as $\left\Vert \mathbf{A}\right\Vert =\max_{\mathbf{%
\boldsymbol{\zeta }\in }R^{s}\mathbf{,||\boldsymbol{\zeta }||=}1}\left\Vert
\mathbf{A\boldsymbol{\zeta }}\right\Vert $, and let $\lambda _{\min }(%
\mathbf{A)}$ and $\lambda _{\max }(\mathbf{A)}$ be the smallest and largest
eigenvalues of $\mathbf{A}$, respectively. For any matrix $\mathbf{A=}\left(
A_{ij}\right) _{i=1,j=1}^{s,t}$, denote $\left\Vert \mathbf{A}\right\Vert
_{\infty }=\max_{1\leq i\leq s}\sum\nolimits_{j=1}^{t}\left\vert
A_{ij}\right\vert $. We introduce the following conditions.

\begin{enumerate}
\item[(C1)] Assume $\left\Vert \mathbf{X}_{j}\right\Vert =\sqrt{n}$, for $%
1\leq j\leq p$, $\lambda _{\min }[(\mathbf{Z},\mathbf{X)}^{\text{T}}(\mathbf{%
\ Z},\mathbf{X)]\geq }C_{1}\left\vert \mathcal{G}_{\min }\right\vert $, and $%
|| \mathbf{X||}_{\infty }\leq C_{2}p$ for some constants $0<C_{1}<\infty $
and $0<C_{2}<\infty $.

\item[(C2)] $p_{\gamma }(t,\lambda )$ is a symmetric function of $t$, and it
is non-decreasing and concave in $t$ for $t$ $\in \lbrack 0,\infty )$. $\rho
(t)$ is a constant for all $t\geq a\lambda $ for some constant $a>0$, and $%
\rho (0)=0$. $\rho ^{\prime }(t)$ exists and is continuous except for a
finite number of $t$ and $\rho ^{\prime }(0+)=1$.

\item[(C3)] The noise vector $\mathbf{\boldsymbol{\epsilon }=(}\epsilon
_{1},\ldots ,\epsilon _{n})^{\text{T}}$ has sub-Gaussian tails such that $P(|%
\mathbf{a}^{\text{T}}\mathbf{\boldsymbol{\epsilon }}|>\mathbf{||a||}x)\leq
2\exp (-c_{1}x^{2})$ for any vector $\mathbf{a\in }R^{n}$ and $x>0$, where $%
0<c_{1}<\infty $.
\end{enumerate}

Conditions (C2)\ and (C3) are common assumptions in
high-dimensional settings. The concave penalties such as MCP and SCAD
satisfy Condition (C2). In the literature, it is commonly assumed that the
smallest eigenvalue of the design matrix is bounded by $C_{1}n$, which may
not hold for $(\mathbf{Z},\mathbf{X)}^{\text{T}}(\mathbf{Z},\mathbf{X)}$.
For instance, by letting $\mathbf{Z}^{\text{T}}\mathbf{X=0}$ and assuming $%
\lambda_{\min}(\mathbf{X}^{\text{T}}\mathbf{X)=}Cn$, we have
\begin{equation*}
\lambda_{\min}[(\mathbf{Z},\mathbf{X)}^{\text{T}}(\mathbf{Z},\mathbf{X)]\geq
}\min \{\lambda_{\min}(\mathbf{D),}\lambda_{\min}(\mathbf{X}^{\text{T}}%
\mathbf{X)\}=}\min (\left\vert \mathcal{G}_{\min}\right\vert ,Cn\mathbf{),}
\end{equation*}%
and $\left\vert \mathcal{G}_{\min}\right\vert \leq n/K$. Therefore, we let
the smallest eigenvalue in Condition (C1) be bounded by $C_{1}\left\vert
\mathcal{G}_{\min}\right\vert $.

When the true group memberships $\mathcal{G}_{1},\ldots ,\mathcal{G}_{K}$
are known, the oracle estimators for $\mathbf{\boldsymbol{\mu }}$ and $%
\mathbf{\boldsymbol{\beta }}$ are
\begin{equation}
(\widehat{\mathbf{\boldsymbol{\mu }}}^{or},\mathbf{\ }\widehat{\mathbf{%
\boldsymbol{\beta }}}^{or})=\arg \min_{\mathbf{\ \boldsymbol{\mu }\in }%
\mathcal{M}_{\mathcal{G}},\mathbf{\boldsymbol{\beta }\in }R^{p}}\frac{1}{2}||%
\mathbf{y-\boldsymbol{\mu }-X\boldsymbol{\beta }||}^{2},  \label{EQ:oracle}
\end{equation}%
and correspondingly, the oracle estimators for the common intercepts ${%
\boldsymbol{\alpha }}$ and the coefficients $\mathbf{\boldsymbol{\beta }}$
are given by
\begin{eqnarray*}
(\widehat{\mathbf{\boldsymbol{\alpha }}}^{or},\mathbf{\ }\widehat{\mathbf{%
\boldsymbol{\beta }}}^{or}) &=&\arg \min_{\mathbf{\boldsymbol{\alpha }\in }%
R^{K},\mathbf{\boldsymbol{\beta }\in }R^{p}}\frac{1}{2}||\mathbf{y-Z%
\boldsymbol{\alpha }-X\boldsymbol{\beta }||}^{2} \\
&=&[(\mathbf{Z},\mathbf{X)}^{\text{T}}(\mathbf{Z},\mathbf{X)]}^{-1}(\mathbf{Z%
},\mathbf{X)}^{\text{T}}\mathbf{y.}
\end{eqnarray*}%
%
Let ${\boldsymbol{\alpha }}^{0}=(\alpha _{k}^{0},k=1,\ldots ,K)^{\text{T}}$,
where $\alpha _{k}^{0}$ is the underlying common intercept for group $%
\mathcal{G}_{k}$. Let $\boldsymbol{\beta }^{0}$ be the underlying regression
coefficient.

\begin{theorem}
\label{THM:norm} Suppose conditions (C1)-(C3) hold. If $K=o(n)$, $p=o(n)$,
and
\begin{equation*}
\left\vert \mathcal{G}_{\min}\right\vert \gg \sqrt{(K+p)n\log n},
\end{equation*}
we have with probability at least $1-2(K+p)n^{-1}$,
\begin{equation}
\left\Vert ((\widehat{\mathbf{\boldsymbol{\mu}}}^{or}-\mathbf{\boldsymbol{\
\mu}}^{0})^{\text{T}},(\widehat{\mathbf{\boldsymbol{\beta}}}^{or}-\mathbf{\
\boldsymbol{\beta}}^{0})^{\text{T}})^{\text{T}}\right\Vert_{\infty}\leq
\phi_{n},  \label{EQ:supnormmubeta}
\end{equation}%
where
\begin{equation}
\phi _{n}=c_{1}^{-1/2}C_{1}^{-1}\sqrt{K+p}\left\vert \mathcal{G}_{\min
}\right\vert ^{-1}\sqrt{n\log n}.  \label{EQ:phn}
\end{equation}%
Moreover, for any vector $\mathbf{a}_{n}\in R^{K+p}$, we have as $%
n\rightarrow \infty $,
\begin{equation}
\sigma_{n}^{-1}(\mathbf{a}_{n}) \mathbf{a}_{n}^{\text{T}} ((\widehat{{%
\boldsymbol{\alpha}}}^{or}-{\boldsymbol{\alpha} }^{0})^{\text{T}},(\widehat{%
\mathbf{\ \boldsymbol{\beta}}}^{or}-\mathbf{\boldsymbol{\beta}}^{0})^{\text{T%
}})^{ \text{T}}\rightarrow N(0,1\mathbf{),}  \label{EQ:normal}
\end{equation}%
where
\begin{equation}
\sigma _{n}(\mathbf{a}_{n})=\sigma \left\{ \mathbf{a}_{n}^{\text{T}}[(%
\mathbf{\ Z},\mathbf{X)}^{\text{T}}(\mathbf{Z},\mathbf{X)]}^{-1}\mathbf{a}%
_{n}\right\}^{1/2}.  \label{EQ:sign}
\end{equation}
\end{theorem}
The proof of this theorem is given in the online supplement.

\textbf{Remark 3.} Since $\left\vert \mathcal{G}_{\min }\right\vert \leq n/K$
, by the condition $\left\vert \mathcal{G}_{\min }\right\vert \gg \sqrt{
(K+p)n\log n}$, $K$ and $p$ must satisfy $K\sqrt{(K+p)}=o(\sqrt{n(\log
n)^{-1}})$, and hence $K=o(n^{1/3}(\log n)^{-1/3})$. By letting $\left\vert
\mathcal{G}_{\min }\right\vert =\delta n/K$ for some constant $0<\delta \leq
1$, the bound (\ref{EQ:supnormmubeta}) is $c_{1}^{-1/2}C_{1}^{-1}\delta
^{-1}K\sqrt{K+p} \sqrt{\log n/n}$. Moreover, when $K$ and $p$ are fixed
numbers, the bound ( \ref{EQ:supnormmubeta}) is $C^{\ast }\sqrt{\log n/n}$
for some constant $0<C^{\ast }<\infty $.

Let
\begin{equation*}
b_{n}=\min_{i\in \mathcal{G}_{k},j\in \mathcal{G}_{k^{\prime }},k\neq
k^{\prime }}|\mu _{i}^{0}-\mu_{j}^{0}|=\min_{k\neq k^{\prime }}|\alpha
_{k}^{0}-\alpha _{k^{\prime }}^{0}|
\end{equation*}
be the minimal difference of the common values between two groups.

\begin{theorem}
\label{THM:selection} Suppose the conditions in Theorem \ref{THM:norm} hold.
If $b_{n}>a\lambda $ and $\lambda \gg \phi _{n}$, where $\phi _{n}$ is given
in (\ref{EQ:phn}), then there exists a local minimizer $(\widehat{\mathbf{%
\boldsymbol{\mu }}}(\lambda )^{\text{T}},\widehat{\mathbf{\boldsymbol{\beta }%
}}(\lambda )^{\text{T}})^{\text{T}}$ of the objective function $Q_{n}(%
\mathbf{\boldsymbol{\mu }}{,\mathbf{\boldsymbol{\beta }};\lambda })$ given
in (\ref{EQ:objective}) satisfying
\begin{equation*}
P\left( (\widehat{\mathbf{\boldsymbol{\mu }}}(\lambda )^{\text{T}},\widehat{%
\mathbf{\boldsymbol{\beta }}}(\lambda )^{\text{T}})^{\text{T}}=((\widehat{%
\mathbf{\boldsymbol{\mu }}}^{or})^{\text{T}},(\mathbf{\boldsymbol{\ }}%
\widehat{\mathbf{\boldsymbol{\beta }}}^{or})^{\text{T}})^{\text{T}}\right)
\rightarrow 1.
\end{equation*}
\end{theorem}
The proof of this theorem is given in the online supplement.

\textbf{Remark 4.} The above result holds given that $b_{n}\gg \phi _{n}$. As
discussed in Remark 3, when $K$ is a finite and fixed number and $\left\vert
\mathcal{G}_{\min }\right\vert =\delta n/K$ for some constant $0<\delta \leq
1$, $b_{n}\gg C^{\ast }\sqrt{\log n/n}$ for some constant $0<C^{\ast
}<\infty $. Moroever, Theorem \ref{THM:selection} shows that the oracle
estimator $((\widehat{\mathbf{\boldsymbol{\mu }}}^{or})^{\text{T}},(\widehat{%
\mathbf{\boldsymbol{\beta }}}^{or})^{\text{T}})^{\text{T}}$ is a local
minimizer $(\widehat{\mathbf{\boldsymbol{\mu }}}(\lambda )^{\text{T}},%
\widehat{\mathbf{\boldsymbol{\beta }}}(\lambda )^{\text{T}})^{\text{T}}$ of
the objective function with probability approaching 1. Let $\widehat{\mathbf{%
\boldsymbol{\alpha }}}(\lambda )$ be the distinct values of $\widehat{%
\mathbf{\boldsymbol{\mu }}}(\lambda )$. Also $\widehat{{\boldsymbol{\alpha }}%
}^{or}$ consists of the distinct values of $\widehat{\mathbf{\boldsymbol{\mu
}}}^{or}$. By the oracle property in Theorem \ref{THM:selection}, we have $%
P\{\widehat{\mathbf{\boldsymbol{\alpha }}}(\lambda )=\widehat{{\boldsymbol{%
\alpha }}}^{or}\}\rightarrow 1$. This result together with the asymptotic
normality given in Theorem \ref{THM:norm} directly leads to the asymptotic
distribution of $(\widehat{\mathbf{\boldsymbol{\alpha }}}(\lambda )^{\text{T}%
},\widehat{\mathbf{\boldsymbol{\beta }}}(\lambda )^{\text{T}})^{\text{T}}$
presented in the following corollary.

\begin{corollary}
\label{COR:distribution}Under the conditions in Theorem \ref{THM:selection},
we have for any vector $\mathbf{a}_{n}\in R^{K+p}$, as $n\rightarrow \infty $
,
\begin{equation*}
\sigma _{n}^{-1}(\mathbf{a}_{n})\mathbf{a}_{n}^{\text{T}}((\widehat{\mathbf{%
\boldsymbol{\alpha }}}(\lambda )-\mathbf{\boldsymbol{\alpha }}^{0})^{\text{T}%
},(\widehat{\mathbf{\boldsymbol{\beta }}}(\lambda )-\mathbf{\boldsymbol{\
\beta }}^{0})^{\text{T}})^{\text{T}}\rightarrow N(0,1\mathbf{),}
\end{equation*}%
with $\sigma _{n}(\mathbf{a}_{n})$ given in (\ref{EQ:sign}). As a result, we
have for any vectors $\mathbf{a}_{n1}\in R^{K}$ and $\mathbf{a}_{n2}\in R^{p}
$, as $n\rightarrow \infty $, $\sigma _{n1}^{-1}(\mathbf{a}_{n1})\mathbf{a}%
_{n1}^{\text{T}}(\widehat{\mathbf{\boldsymbol{\alpha }}}(\lambda )-\mathbf{%
\boldsymbol{\alpha }}^{0})\rightarrow N(0,1\mathbf{)}$ and $\sigma
_{n2}^{-1}(\mathbf{a}_{n2})\mathbf{a}_{n2}^{\text{T}}(\widehat{\mathbf{%
\boldsymbol{\beta }}}(\lambda )-\mathbf{\boldsymbol{\beta }}^{0})\rightarrow
N(0,1\mathbf{)}$, where
\begin{eqnarray*}
\sigma _{n1}(\mathbf{a}_{n1}) &=&\sigma \left[ \mathbf{a}_{n1}^{\text{T}}\{%
\mathbf{Z}^{\text{T}}\mathbf{Z-(Z}^{\text{T}}\mathbf{X)(X}^{\text{T}}\mathbf{%
\ X)}^{-1}\mathbf{(X}^{\text{T}}\mathbf{Z)\}}^{-1}\mathbf{a}_{n1}\right]
^{1/2}, \\
\sigma _{n2}(\mathbf{a}_{n2}) &=&\sigma \left[ \mathbf{a}_{n2}^{\text{T}}\{%
\mathbf{X}^{\text{T}}\mathbf{X-(X}^{\text{T}}\mathbf{Z)(Z}^{\text{T}}\mathbf{%
\ Z)}^{-1}\mathbf{(Z}^{\text{T}}\mathbf{X)\}}^{-1}\mathbf{a}_{n2}\right]
^{1/2}.
\end{eqnarray*}
\end{corollary}

\textbf{Remark 5.} The asymptotic distribution of the penalized estimators
provides a theoretical justification for further conducting statistical
inference about subgrouping. By the results in Corollary \ref%
{COR:distribution}, for given $\mathbf{a}_{n1}\in R^{K}$ and $\mathbf{a}
_{n2}\in R^{p}$, $100(1-\alpha )\%$ confidence intervals for $\mathbf{a}
_{n1}^{\text{T}}\mathbf{\boldsymbol{\mu }}^{0}$ and $\mathbf{a}_{n2}^{\text{%
T }}\mathbf{\boldsymbol{\beta }}^{0}$ are given as $\mathbf{a}_{n1}^{\text{T}%
} \widehat{\mathbf{\boldsymbol{\alpha }}}(\lambda )\pm z_{\alpha /2}\widehat{
\sigma }_{n1}(\mathbf{a}_{n1})$ and $\mathbf{a}_{n2}^{\text{T}}\widehat{
\mathbf{\boldsymbol{\beta }}}(\lambda )\pm z_{\alpha /2}\widehat{\sigma }
_{n2}(\mathbf{a}_{n2})$, respectively, where $z_{\alpha /2}$ is the $%
(1-\alpha /2)100$ percentile of the standard normal, and $\widehat{\sigma }
_{n1}(\mathbf{a}_{n1})$ and $\widehat{\sigma }_{n2}(\mathbf{a}_{n2})$ are
estimates of $\sigma _{n1}(\mathbf{a}_{n1})$ and $\sigma _{n2}(\mathbf{a}
_{n2})$ with $\sigma ^{2}$ estimated by $\widehat{\sigma }^{2}=(n-\widehat{K}
-p)^{-1}\sum\nolimits_{i=1}^{n}(y_{i}-\widehat{\mu }_{i}-\mathbf{x}_{i}^{
\text{T}}\widehat{\mathbf{\boldsymbol{\beta }}}\mathbf{)}^{2}$, where $\widehat{K}$
is the number of distinct values in $\widehat{\mathbf{\boldsymbol{\mu }}}%
(\lambda )$.

\section{Simulation studies\label{SEC:examples}}

In this section, we conduct simulation experiments to investigate the
numerical performance of our proposed estimators.

We use the modified Bayesian Information Criterion (BIC) (\cite%
{wang.li.tsai:2007}) for high-dimensional data settings to select the tuning
parameter by minimizing%
\begin{equation}
\text{BIC}=\log [\sum\nolimits_{i=1}^{n}(y_{i}-\widehat{\mu }_{i}-\mathbf{x}%
_{i}^{\text{T}}\widehat{\mathbf{\boldsymbol{\beta} }})^{2}/n]+C_{n}\frac{%
\log n}{n}(\widehat{K}+p),  \label{EQ:BIC}
\end{equation}%
where $C_{n}$ is a positive number which can depend on $n$. When $C_{n}=1$,
the modified BIC reduces to the traditional BIC (\cite{schwarz:1978}). \cite%
{wang.leng:2009} used $C_{n}=\log (\log (d))$ in their simulation study when
the number of predictors which is $d$ diverges with sample size. In this
paper, we adopt the same strategy and let $C_{n}=c\log (\log (d))$, where $%
d=n+p$ and $c$ is a positive constant. In our analysis, we select $\lambda $
by minimizing the modified BIC and use a fixed value for $\vartheta $ and $%
\gamma $.

\textbf{Example 1.} We simulate data from the model
\begin{equation}
y_{i}=\mu _{i}+\mathbf{x}_{i}^{\text{T}}\mathbf{\boldsymbol{\beta }+}%
\epsilon _{i},i=1,\ldots ,n,  \label{EQ:DGP}
\end{equation}%
where $\mathbf{x}_{i}=(x_{i1},\ldots ,x_{i5})^{\text{T}}$ are generated from
the multivariate normal distribution with mean $0$, variance $1$ and an
exchangeable correlation $\rho =0.3$ and the error terms $\epsilon _{i}$ are
from independent $N(0,0.5^{2})$. We simulate $\mathbf{\boldsymbol{\beta }=(}%
\boldsymbol{\beta }_{1},\ldots ,\boldsymbol{\beta }_{5})^{\text{T}}$ from
independent Uniform$[0.5,1]$. We generate $\mu _{i}$ from two different
values $-\alpha $ and $\alpha $ with equal probabilities, i.e., we generate
them from the distribution: $p(\mu _{i}=-\alpha )=p(\mu _{i}=\alpha )=1/2$,
so that there are two intercepts $\alpha _{1}=-\alpha $ and $\alpha
_{2}=\alpha $. In our simulation studies, we take different values of $%
\alpha $ for illustration of our proposed method. It is noteworthy that for
smaller value of $\alpha $, it is more difficult to identify the two groups.

In our analysis, we choose to fix $\vartheta =1$ and $\gamma =3$. We compare
the performance of the estimators with the ADMM algorithm by using
the two concave penalties (MCP and SCAD) and using a weighted $L_{1}$
penalty
\begin{equation*}
p_{\gamma }(|\mu_i-\mu_j|,\lambda )=\lambda \omega _{ij}|\mu _{i}-\mu_j|,
\end{equation*}
which requires specification of the weights $\omega_{ij}$. As discussed in
\cite{chi.lange:2014}, the choice of the weights can dramatically affect the
quality of the results in cluster analysis. In the regression context such as in
our study, it is even more challenging to select the weights.
%
%

For the $L_{1}$ penalty, we let the weight be $\omega _{ij}=\exp (-\phi
(y_{i}-y_{j})^{2})$ which is a Gaussian kernel defined on the distance of
two points. The constant $\phi $ is nonnegative. When $\phi =0$, it
corresponds to the Lasso penalty. Note that it is unclear what weights we
need to apply to obtain optimal results. We here use the Gaussian kernel as
the weight to illustrate this point by using different values for $\phi $.

{\normalsize
\begin{figure}[tbp]
\caption{Solution paths for the means $(\protect\mu _{1},\ldots ,\protect\mu %
_{n})$ against $\protect\lambda $ values by using MCP, SCAD and $L_{1}$
penalties, respectively, in Example 1.}
\label{FIG:path}{\normalsize \vspace{-0.5cm}  }
\par
\begin{center}
{\normalsize $%
\begin{array}{cc}
\includegraphics[width=2.7in]{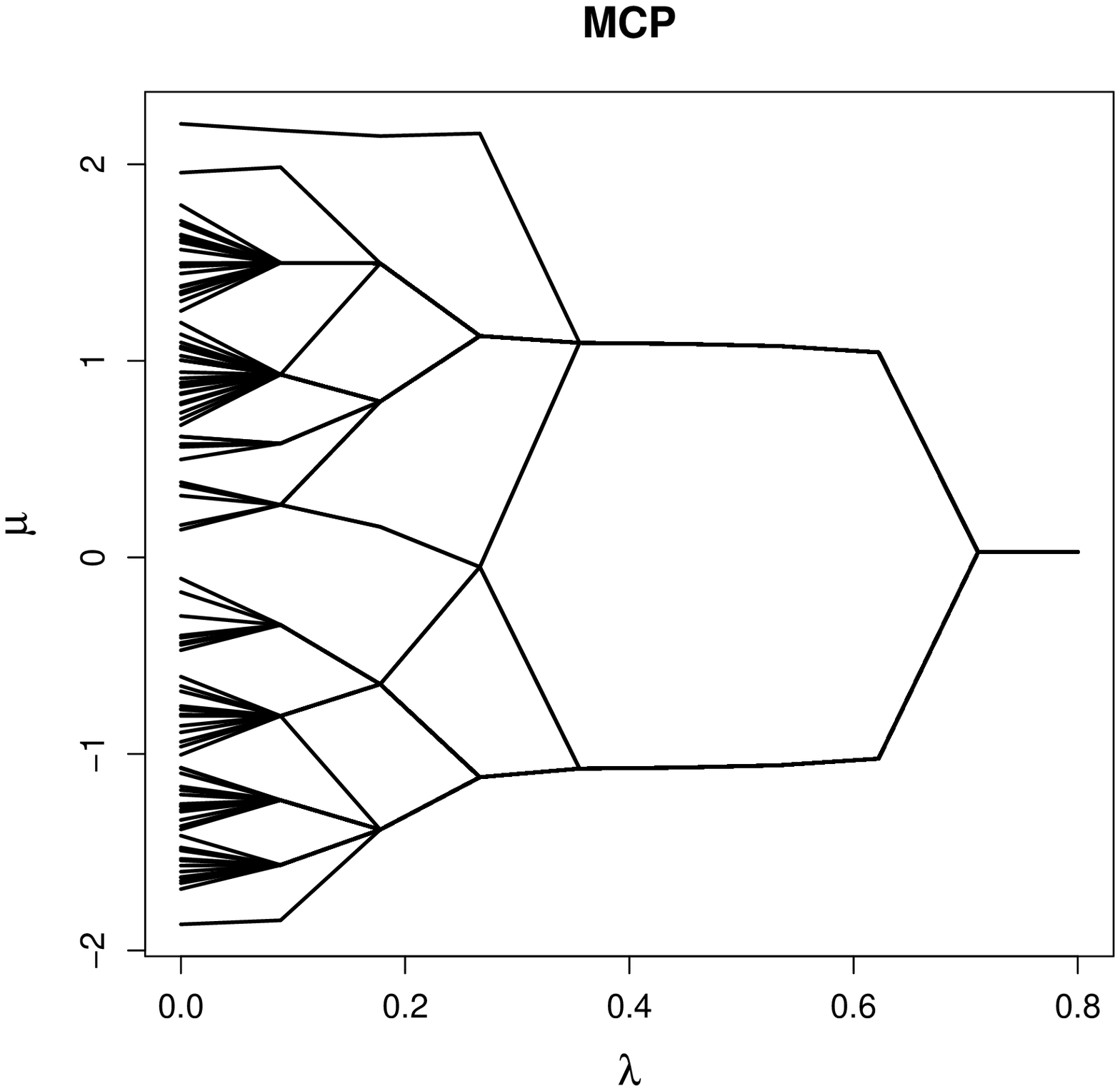} & %
\includegraphics[width=2.7in]{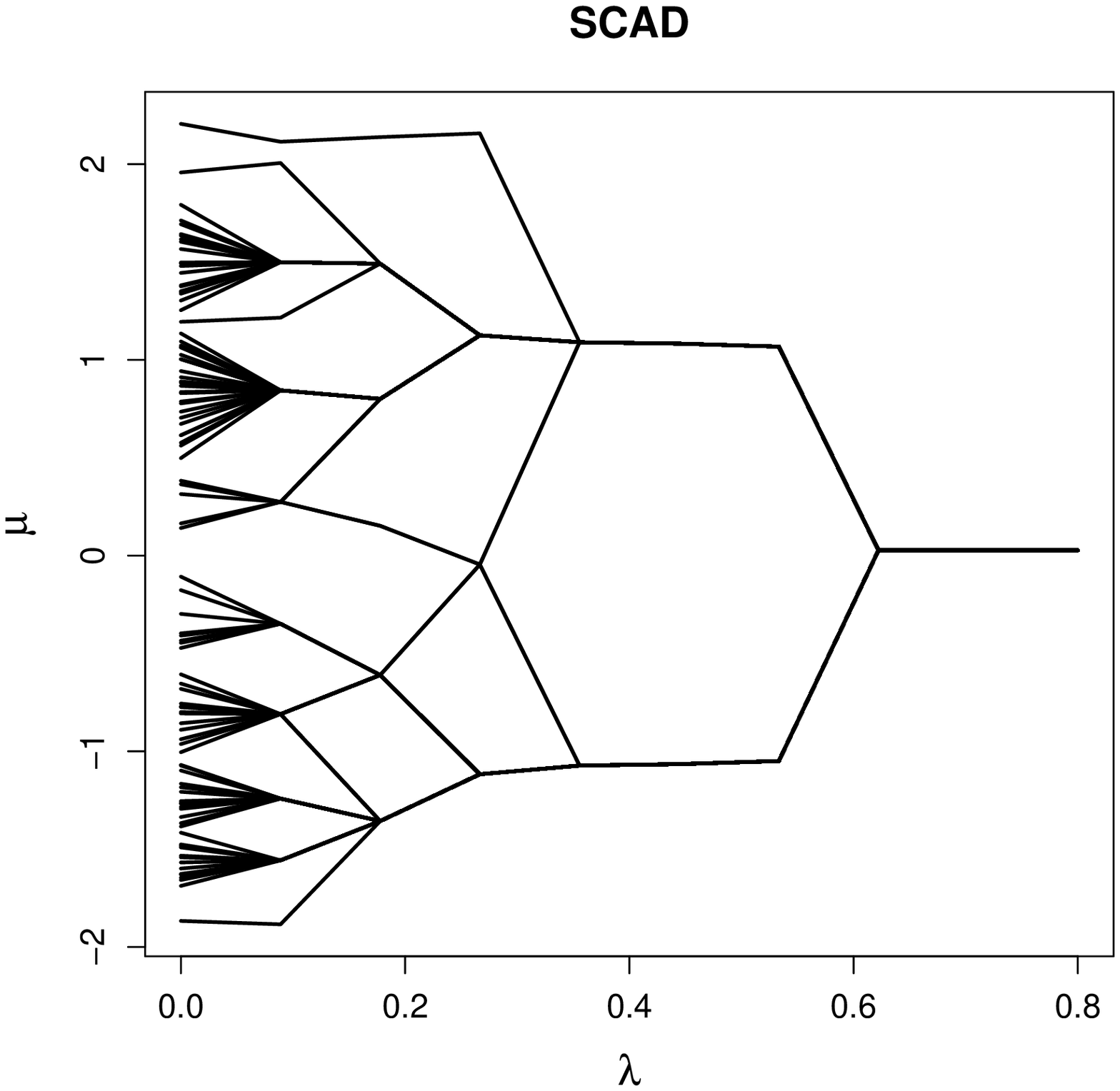} \\
\includegraphics[width=2.7in]{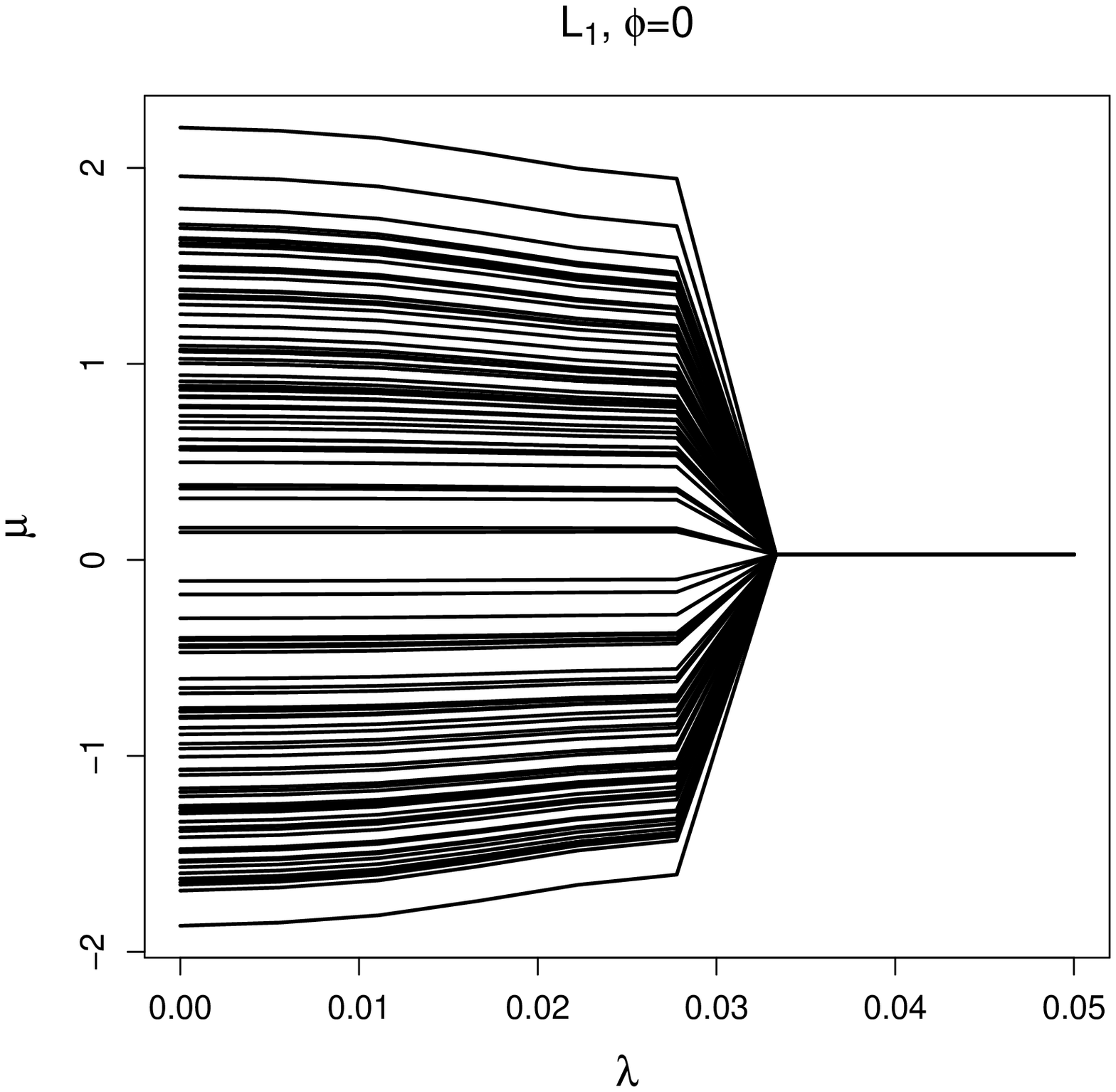} & %
\includegraphics[width=2.7in]{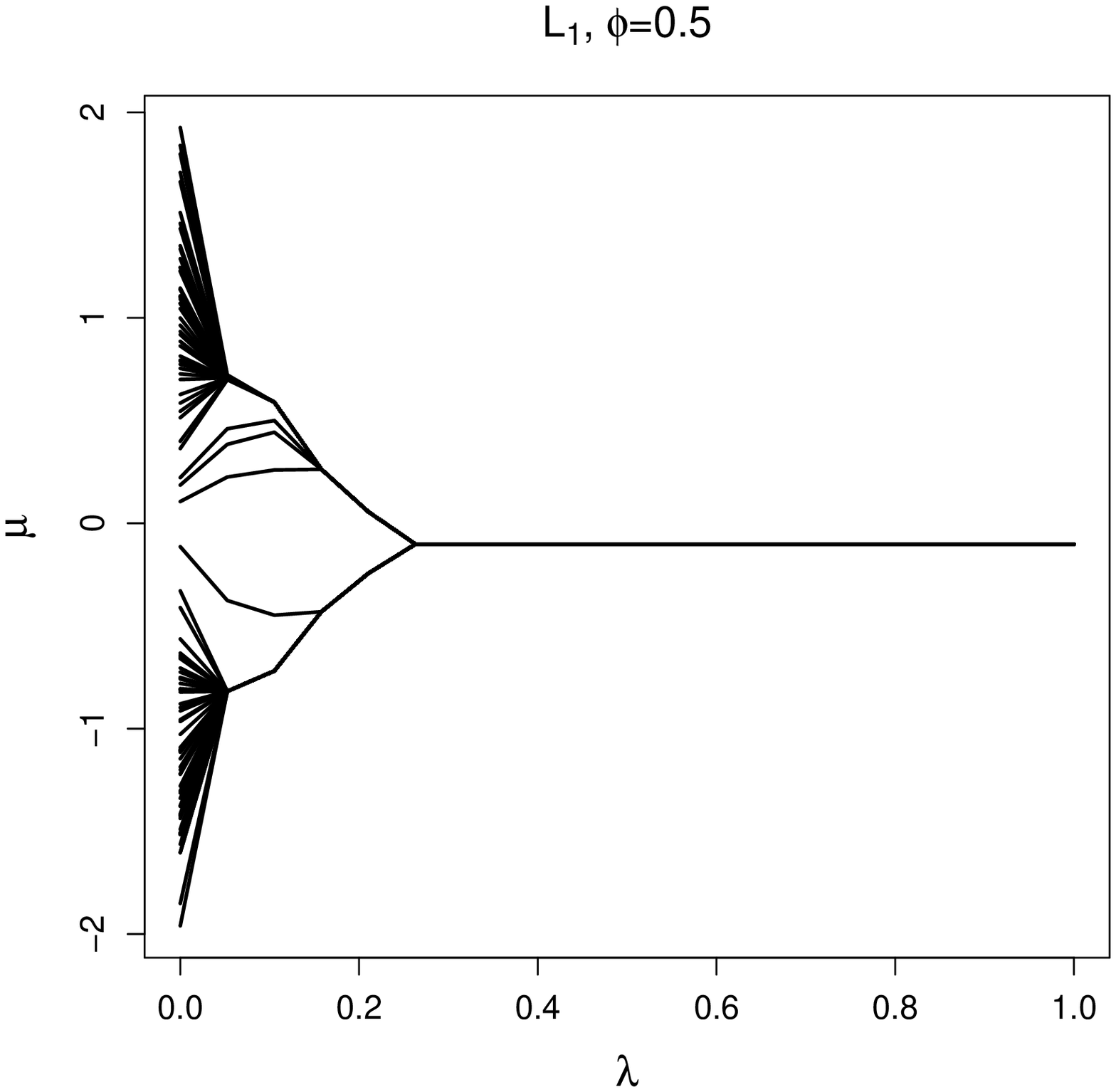} \\
\includegraphics[width=2.7in]{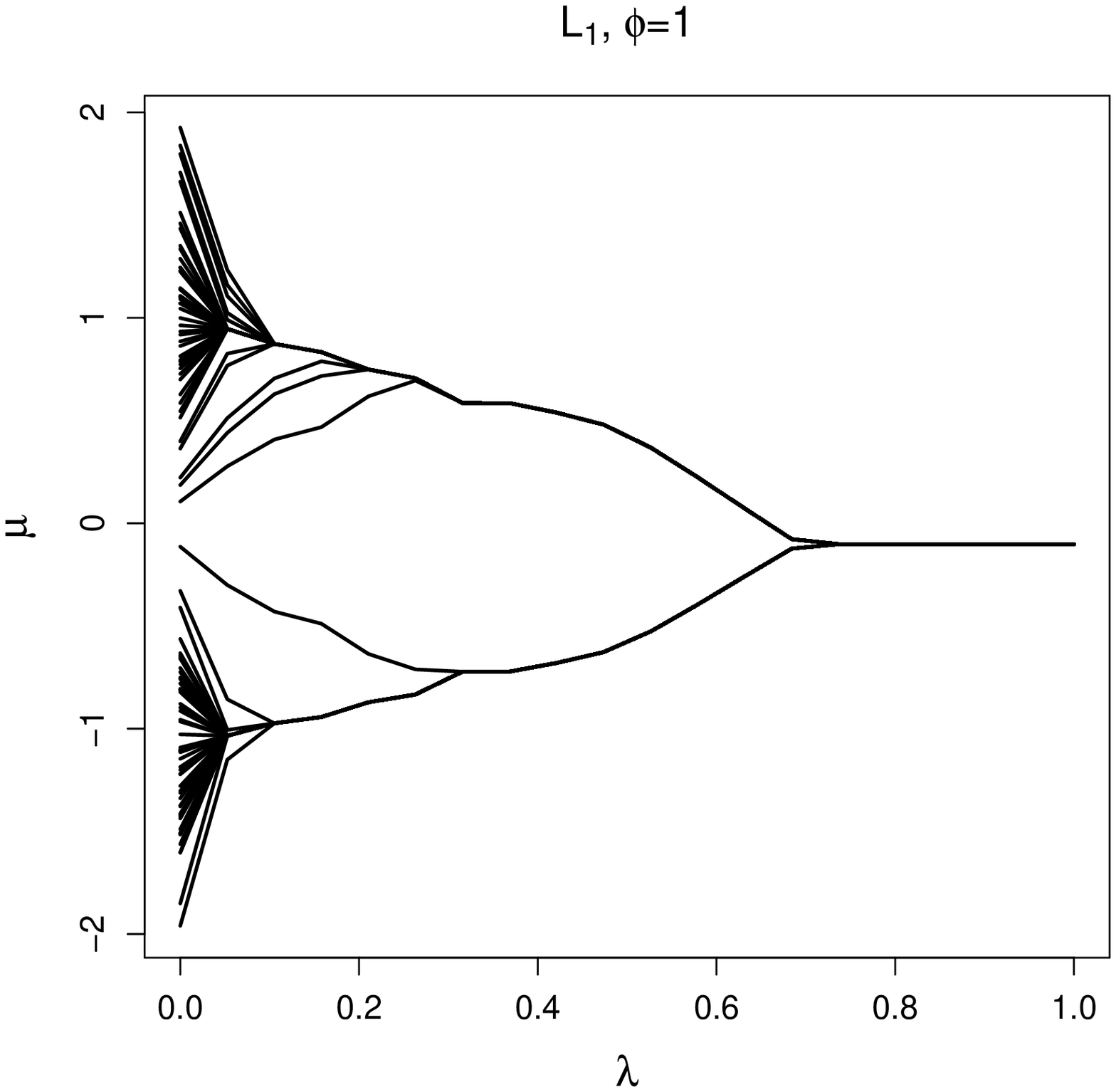} & %
\includegraphics[width=2.7in]{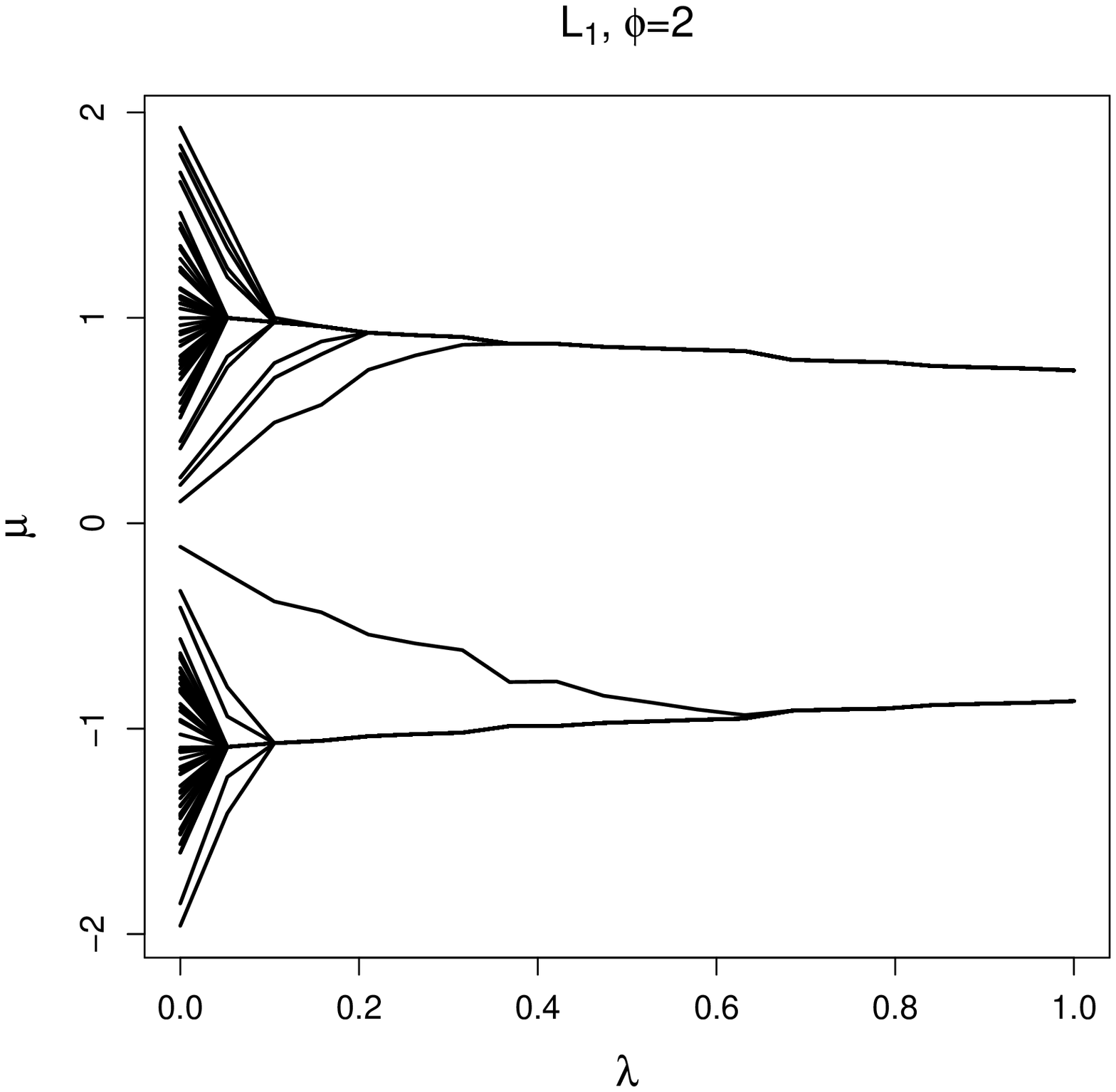}%
\end{array}
$  }
\end{center}
\end{figure}
}
Figure \ref{FIG:path} displays the solution paths for the means $(\mu
_{1},\ldots ,\mu _{n})$ against $\lambda $ values by using MCP and SCAD, and
the $L_{1}$ penalties with $\phi =0,0.5,1,2$, respectively, based on one
sample with $n=100$ and $\alpha =1$. We observe that the MCP and SCAD have
similar solution paths as shown in Figure \ref{FIG:path}. For these two
penalties, the estimated values for $\mathbf{\boldsymbol{\mu }}$ converge to
two different values around $-1$ and $1$ which are the true values for the
intercepts of the two groups, when $\lambda $ reaches certain value (around $%
0.38$ for both MCP and SCAD). They eventually converge to one value when $%
\lambda $ exceeds $0.6$. The $L_{1}$ penalty, however, shows a different
solution path from MCP and SCAD, and the solution paths look quite
differently for different values of $\phi $, so that the choice of weights
can dramatically affect the estimation results. When $\phi =0$ which is the
LASSO penalty, we see that the estimated values for $\mu _{i}$'s converge quickly as the $\lambda $ value increases until they converge to a common
point around $0$ when $\lambda $ reaches $0.035$. As a result, it cannot
effectively identify the groups of the $\mathbf{\boldsymbol{\mu }}$ value.
By looking at the plots for $\phi =0.5,1,2$, we observe that as the $\phi $
value becomes larger, the estimated values converge to one point more slowly.%

Next we conduct the simulations by selecting $\lambda $ via minimizing the
modified BIC\ given in (\ref{EQ:BIC}). Recall that we let $C_{n}=c\log (\log
(n+p))$, where $n+p$ is the number of components in $\mathbf{\boldsymbol{\mu
}}$ and $\mathbf{\boldsymbol{\beta }}$, and $c$ is a positive constant. We
use different $c$ values by letting $c=5,10$ in our estimation procedure. We
consider different values for $\alpha $ by letting $\alpha =1,1.5,2$, so
that the difference of the true common values between the two groups varies
from $2$ to $4$. Table \ref{TAB:K} reports the mean, the median and standard
error (s.e.) of the estimated number of groups $\widehat{K}$ by the MCP,
SCAD and $L_{1}$ methods with $\phi =1$ and $2$ based on $100$ simulation
realizations with $n=100$. Moreover, to study the estimation accuracy, in
Table \ref{TAB:beta} we report the average value and the standard error
shown in the parentheses of the square root of the mean squared errors (MSE)
for the estimated values of $\mathbf{\boldsymbol{\mu }}$ and $\mathbf{%
\boldsymbol{\ \beta }}$ for the MCP, SCAD and $L_{1}$ estimators and the
oracle estimator given in (\ref{EQ:oracle}). The square roots of the MSE
for $\mathbf{\ \boldsymbol{\mu }}$ and $\mathbf{\boldsymbol{\beta }}$ are,
respectively, defined as $\left\Vert \widehat{\mathbf{\boldsymbol{\mu }}}-%
\mathbf{\ \boldsymbol{\mu }}\right\Vert /\sqrt{n}$ and $||\widehat{\mathbf{%
\boldsymbol{\ \beta }}}-\mathbf{\boldsymbol{\beta }||}/\sqrt{p}$ for each
realization.{\normalsize \
\begin{table}[tbph]
\caption{The mean, median and standard error (s.e.) of $\protect\widehat{K}$
by the MCP, SCAD and $L_{1}$ methods with $\protect\phi =1.0$ and $2.0$
based on 100 realizations with $n=100$ in Example 1.}
\label{TAB:K}
\begin{center}
{\normalsize
\begin{tabular*}{0.95\textwidth}{l|l|@{\extracolsep{\fill}}ccc|ccc|ccc}
\cline{1-11}
$c$ & $\boldsymbol{\alpha} $ &  & $1.0$ &  &  & $1.5$ &  &  & $2.0$ &  \\
\cline{1-11}
&  & mean & median & s.e. & mean & median & s.e. & mean & median & s.e. \\
\cline{1-11}
& MCP & 2.57 & 2.00 & 0.90 & 2.41 & 2.00 & 0.93 & 2.10 & 2.00 & 0.44 \\
$5.0$ & SCAD & 2.58 & 2.00 & 0.96 & 2.37 & 2.00 & 0.90 & 2.18 & 2.00 & 0.63
\\
& L$_{1}(\phi =1.0)$ & 1.76 & 1.00 & 0.99 & 2.71 & 3.00 & 0.88 & 2.50 & 2.00
& 0.82 \\
& L$_{1}(\phi =2.0)$ & 3.03 & 3.00 & 1.16 & 3.13 & 3.00 & 1.19 & 3.25 & 3.00
& 1.00 \\\cline{1-11}
& MCP & 2.10 & 2.00 & 0.33 & 2.04 & 2.00 & 0.20 & 2.01 & 2.00 & 0.11 \\
$10.0$ & SCAD & 2.11 & 2.00 & 0.35 & 2.04 & 2.00 & 0.20 & 2.02 & 2.00 & 0.14
\\
& L$_{1}(\phi =1.0)$ & 1.40 & 1.00 & 0.65 & 5.10 & 4.00 & 3.00 & 3.75 & 3.00
& 1.60 \\
& L$_{1}(\phi =2.0)$ & 2.29 & 2.00 & 0.78 & 3.03 & 3.00 & 1.02 & 3.25 & 3.00
& 1.00 \\ \cline{1-11}
\end{tabular*}
}
\end{center}
\end{table}
\begin{table}[tbph]
\caption{The mean and standard error (s.e.) shown in parentheses of the
square root of the MSE for the estimated values of $\mathbf{\boldsymbol{%
\protect\mu }}$ and $\mathbf{\boldsymbol{\protect\beta }}$ for the MCP, SCAD
and $L_{1}$ penalty estimators and the oracle estimators with $\protect\phi %
=1.0$ and $2.0$ based on 100 realizations with $n=100$ in Example 1.}
\label{TAB:beta}
\begin{center}
{\normalsize
\begin{tabular*}{0.95\textwidth}{l|l|@{\extracolsep{\fill}}ccc|ccc}
\hline
&  & \multicolumn{3}{c|}{$\mathbf{\boldsymbol{\mu }}$} & \multicolumn{3}{c}{$%
\mathbf{\boldsymbol{\beta }}$} \\ \hline
$c$ & $\boldsymbol{\alpha} $ & $1.0$ & $1.5$ & $2.0$ & $1.0$ & $1.5$ & $2.0$
\\ \hline
& MCP & 0.409 & 0.246 & 0.132 & 0.043 & 0.076 & 0.062 \\
&  & (0.108) & ( 0.192) & (0.151) & (0.034) & (0.045) & (0.038) \\
$5.0$ & SCAD & 0.414 & 0.240 & 0.158 & 0.091 & 0.075 & 0.065 \\
&  & (0.116) & (0.190) & (0.168) & (0.036) & (0.044) & (0.040) \\
& L$_{1}(\phi =1.0)$ & 0.874 & 0.370 & 0.185 & 0.118 & 0.084 & 0.066 \\
&  & ( 0.202) & (0.237) & (0.173) & (0.040) & (0.047) & (0.036) \\
& L$_{1}(\phi =2.0)$ & 0.637 & 0.274 & 0.167 & 0.106 & 0.076 & 0.064 \\
&  & (0.226) & (0.180) & (0.153) & ( 0.040) & (0.041) & (0.035) \\ \hline
& MCP & 0.407 & 0.230 & 0.154 & 0.086 & 0.069 & 0.062 \\
&  & (0.139) & (0.178) & (0.164) & (0.035) & (0.034) & (0.030) \\
$10.0$ & SCAD & 0.409 & 0.234 & 0.155 & 0.086 & 0.069 & 0.061 \\
&  & (0.138) & (0.178) & (0.163) & (0.034) & (0.035) & (0.030) \\
& L$_{1}(\phi =1.0)$ & 0.946 & 0.265 & 0.203 & 0.121 & 0.075 & 0.069 \\
&  & ( 0.138) & (0.142) & (0.169) & ( 0.039) & (0.038) & (0.038) \\
& L$_{1}(\phi =2.0)$ & 0.769 & 0.287 & 0.167 & 0.113 & 0.078 & 0.064 \\
&  & (0.215) & (0.210) & (0.153) & ( 0.039) & (0.046) & ( 0.035) \\ \hline
\end{tabular*}
}
\end{center}
\end{table}
}

In Table \ref{TAB:K}, for both MCP and SCAD methods we observe that the
median value of $\widehat{K}$ among the 100 replications is $2$ for all
cases, which is the true number of groups in our model, and the mean values
are close to $2$ for different values of $\alpha$. For larger value of $%
\alpha$, it is easier to detect the subgroups, so that correspondingly we
observe that the mean values of $\widehat{K}$ are closer to $2$ for larger $%
\alpha$. The MCP and SCAD can identify the groups for both values of $c$,
although they perform better with $c=10$ by having smaller standard errors.
The $L_{1}$ penalties with both values for $\phi$ in general have worse
performance than the MCP and SCAD penalties. They have mean and
median values for $\widehat{K}$ further away from $2$ and larger
standard errors. Moreover, the performance of the $L_{1}$ penalty is not
stable. The $L_{1}$ penalty with $\phi=1$ tends to select less than two
groups for $\alpha =1.0$ and more than two groups
for $\alpha =1.5, 2.0$, while the $L_{1}$ penalty with $\phi =2$ tends to
select more groups in general. For $\alpha =1.5$ and $2.0$ and $c=5.0$, the $%
L_{1}$ penalty with $\phi =1$ performs better than the $L_{1}$ penalty with $%
\phi =2$ by having $\widehat{K}$ values closer to two and smaller standard
errors, but for other cases the $L_{1}$ penalty with $\phi=2$ seems to
perform better. Thus, we see that different weights applied to the $L_{1}$
penalty may significantly affect the performance of the resulting estimator,
and there is no clear rule on what weight to be used in the general situation.
Table \ref{TAB:beta} shows that the MCP\ and SCAD methods have smaller MSE
values than the $L_{1}$ penalty methods in general since they have more
accurate selection results and produce less biased estimates.

To evaluate the asymptotic normality established in Corollary \ref%
{COR:distribution}, Table \ref{TAB:bias} lists the empirical bias (Bias) for
the estimates of the two intercepts $\alpha _{1}$ and $\alpha _{2}$, and it
also presents the average asymptotic standard error (ASE) calculated
according to Corollary \ref{COR:distribution} and the empirical standard error
(ESE) based on 100 replications for the MCP and SCAD methods with $c=10$ as
well as the oracle estimator (ORACLE). The biases are around zero for all
cases. Moreover, we observe that the asymptotic standard errors for the MCP
and SCAD methods are similar to those for the ORACLE estimator. This result
supports our asymptotic normality result in Corollary \ref{COR:distribution}.%
{\normalsize \
\begin{table}[tbph]
\caption{The empirical bias (Bias) for the estimates of $\protect\alpha _{1}$
and $\protect\alpha _{2}$, and the average asymptotic standard error (ASE)
calculated according to Corollary \protect\ref{COR:distribution} and the
empirical standard error (ESE) based on 100 replications for the MCP\ and SCAD
methods and the oracle estimator (ORACLE) with $c=10$ in Example 1.}
\label{TAB:bias}
\begin{center}
{\normalsize
\begin{tabular*}{0.95\textwidth}{l|c|@{\extracolsep{\fill}}cccccc}
\hline
&  & \multicolumn{2}{c}{$\alpha =1.0$} & \multicolumn{2}{c}{$\alpha =1.5$} &
\multicolumn{2}{c}{$\alpha =2.0$} \\ \hline
&  & $\alpha _{1}$ & $\alpha _{2}$ & $\alpha _{1}$ & $\alpha _{2}$ & $\alpha
_{1}$ & $\alpha _{2}$ \\ \hline
& Bias & 0.037 & -0.066 & 0.031 & -0.055 & 0.065 & -0.083 \\
MCP & ASE & 0.071 & 0.070 & 0.072 & 0.071 & 0.075 & 0.074 \\
& ESE & 0.104 & 0.117 & 0.085 & 0.092 & 0.085 & 0.092 \\ \hline
& Bias & 0.040 & 0.069 & 0.036 & -0.060 & 0.067 & -0.087 \\
SCAD & ASE & 0.071 & 0.070 & 0.072 & 0.071 & 0.075 & 0.074 \\
& ESE & 0.103 & 0.119 & 0.085 & 0.094 & 0.084 & 0.094 \\ \hline
& Bias & -0.009 & -0.005 & -0.010 & -0.005 & -0.010 & -0.005 \\
ORACLE & ASE & 0.072 & 0.072 & 0.072 & 0.072 & 0.072 & 0.072 \\
& ESE & 0.070 & 0.067 & 0.070 & 0.067 & 0.070 & 0.067 \\ \hline
\end{tabular*}
}
\end{center}
\end{table}
}

Lastly, we conduct inferences on the difference between groups. Table \ref%
{TAB:pvalue} presents the average p-values for testing $\mathcal{H}%
_{0}:\alpha _{1}=\alpha _{2}$ based on the 100 simulation realizations. We
use $\sigma _{n1}(\mathbf{a})^{-1}(\widehat{\alpha }_{1}(\lambda )-\widehat{%
\alpha }_{2}(\lambda ))$, $\mathbf{a=(}1,-1)$, as the test statistic which
has the asymptotic normal distribution given in Corollary \ref%
{COR:distribution}, and the estimates $\widehat{\alpha }_{1}(\lambda )$ and $%
\widehat{\alpha }_{2}(\lambda )$ are obtained by the MCP and SCAD methods
with $c=10$. We obtain the p-values close to zero for all cases, so that the
difference between the groups is further confirmed by the inference
procedure.{\normalsize \
\begin{table}[tbph]
\caption{{The average p-values for testing }$\mathcal{H}_{0}:\protect\alpha %
_{1}=\protect\alpha _{2}$ based on the 100 simulation realizations with the
estimates $\protect\widehat{\protect\alpha }_{1}(\protect\lambda )$ and $%
\protect\widehat{\protect\alpha }_{2}(\protect\lambda )$ obtained by the MCP
and SCAD methods with $c=10$ in Example 1.}
\label{TAB:pvalue}
\begin{center}
{\normalsize
\begin{tabular*}{0.95\textwidth}{l|@{\extracolsep{\fill}}ccc}
\hline
$\boldsymbol{\alpha} $ & \multicolumn{1}{c}{$1.0$} & $1.5$ & $2.0$ \\ \hline
MCP & $<0.001$ & $<0.001$ & $<0.001$ \\ \hline
SCAD & $<0.001$ & $<0.001$ & $<0.001$ \\ \hline
\end{tabular*}
}
\end{center}
\end{table}
}

\textbf{Example 2.} We simulate data from model (\ref{EQ:DGP}) with the
predictors, the error terms and the coefficients $\mathbf{\boldsymbol{\beta }%
}$ generated from the same distributions as given in Example 1. We simulate $%
\mu _{i}$ from three different values $-2$, $0$, $2$ with equal
probabilities. We use the modified BIC to select the tuning parameter $%
\lambda $ by letting $C_{n}=5\log (\log (n+p))$. Figure \ref{FIG:boxplot}
shows the boxplots of the estimated number of subgroups $\widehat{K}$ and
the square root of the MSE for the estimated values of $\mathbf{\boldsymbol{%
\ \mu }}$ and $\mathbf{\boldsymbol{\beta }}$, respectively, by using MCP,
SCAD and $L_{1}$ penalty with $\phi =1,2$ methods based on $100$ simulation
realization with $n=100$. In the first plot, we observe that for the MCP and
SCAD methods, the median value for $\widehat{K}$ is $3$, which is the true
number of groups in our model. For some replications, they select more
groups than three. For the $L_{1}$ penalty with $\phi =1$, the median value
for $\widehat{K}$ is $3$ as well. However, some replications have more than
three and others have less than three for the $\widehat{K}$ value. Moreover,
for this example, the $L_{1}$ penalty with $\phi =2$ tends to select more
groups in all replications. The other two plots show that the MCP and SCAD
have much smaller MSE values than the two $L_{1}$ penalty methods.

{\normalsize
\begin{figure}[t!]
\caption{Boxplots of the estimated number of subgroups $\protect\widehat{K}$
and the square root of the MSE for the estimated values of $\mathbf{%
\boldsymbol{\protect\mu }}$ and $\mathbf{\boldsymbol{\protect\beta }}$,
respectively, by using MCP, SCAD and $L_{1}$ with $\protect\phi =1,2$
methods based on $100$ simulation realizations with $n=100$ in Example 2.}
\label{FIG:boxplot}{\normalsize \vspace{-0.3cm}  }
\par
\begin{center}
{\normalsize $%
\begin{array}{cc}
\includegraphics[width=3in]{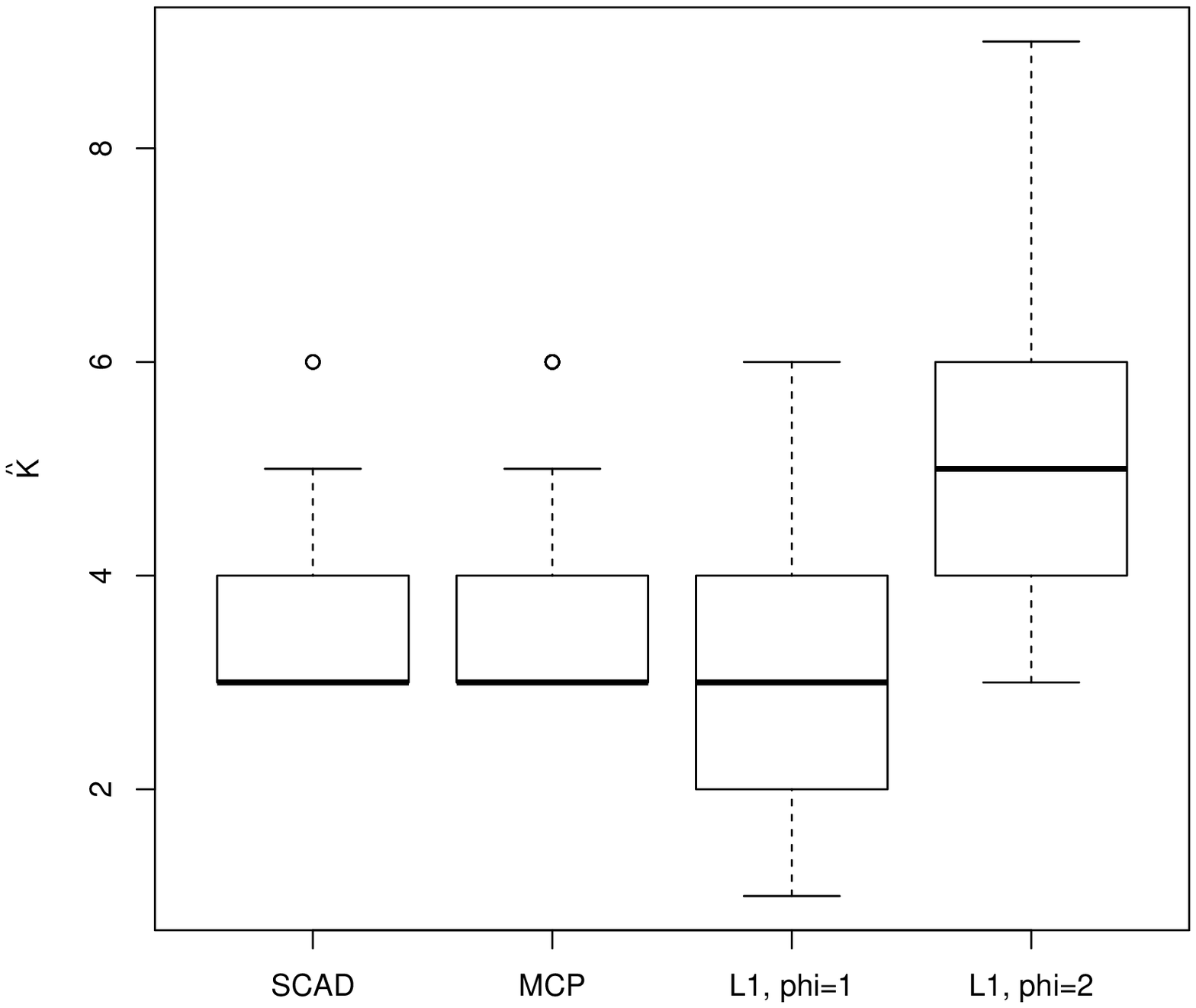} & %
\includegraphics[width=3in]{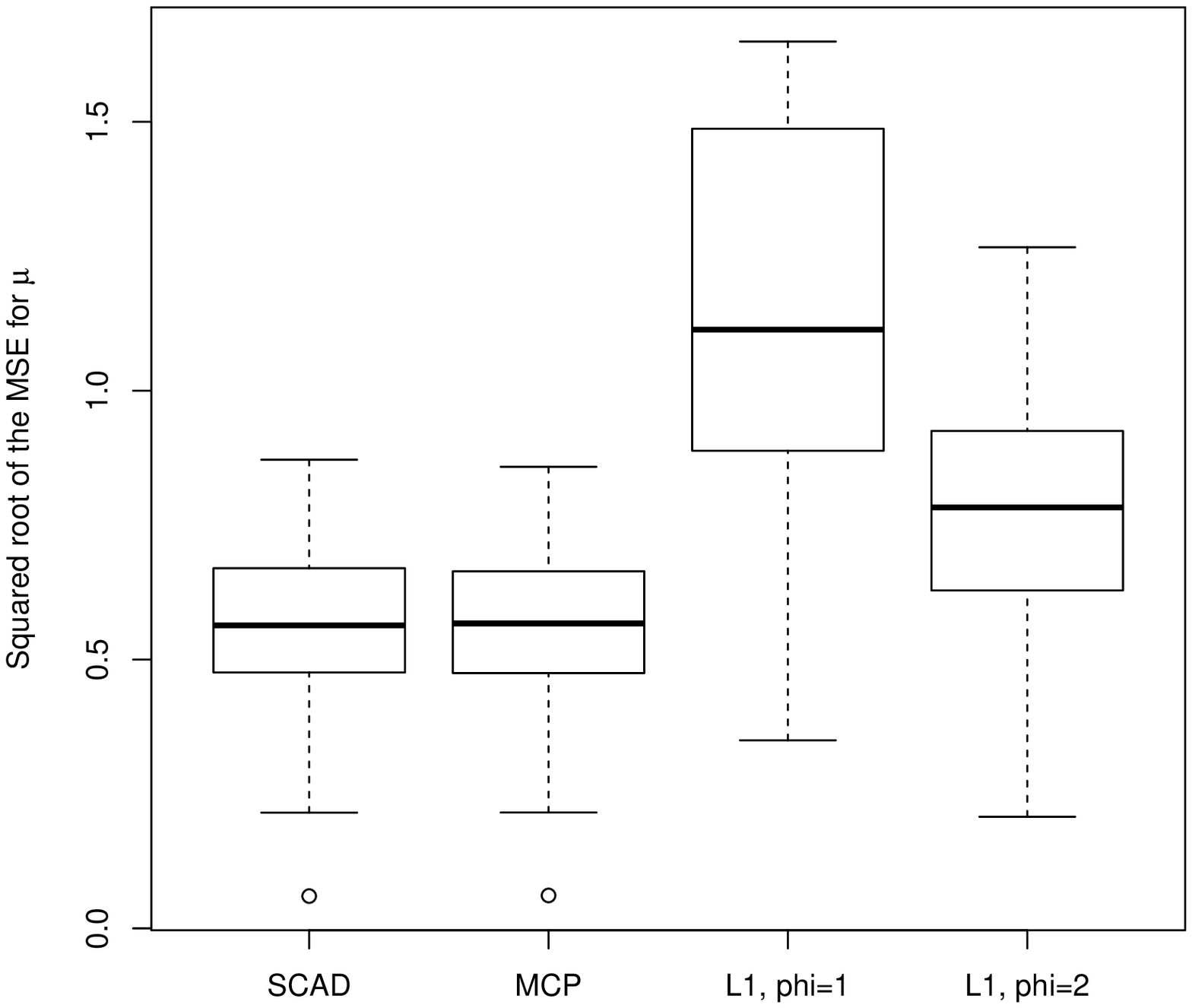} \\
\includegraphics[width=3in]{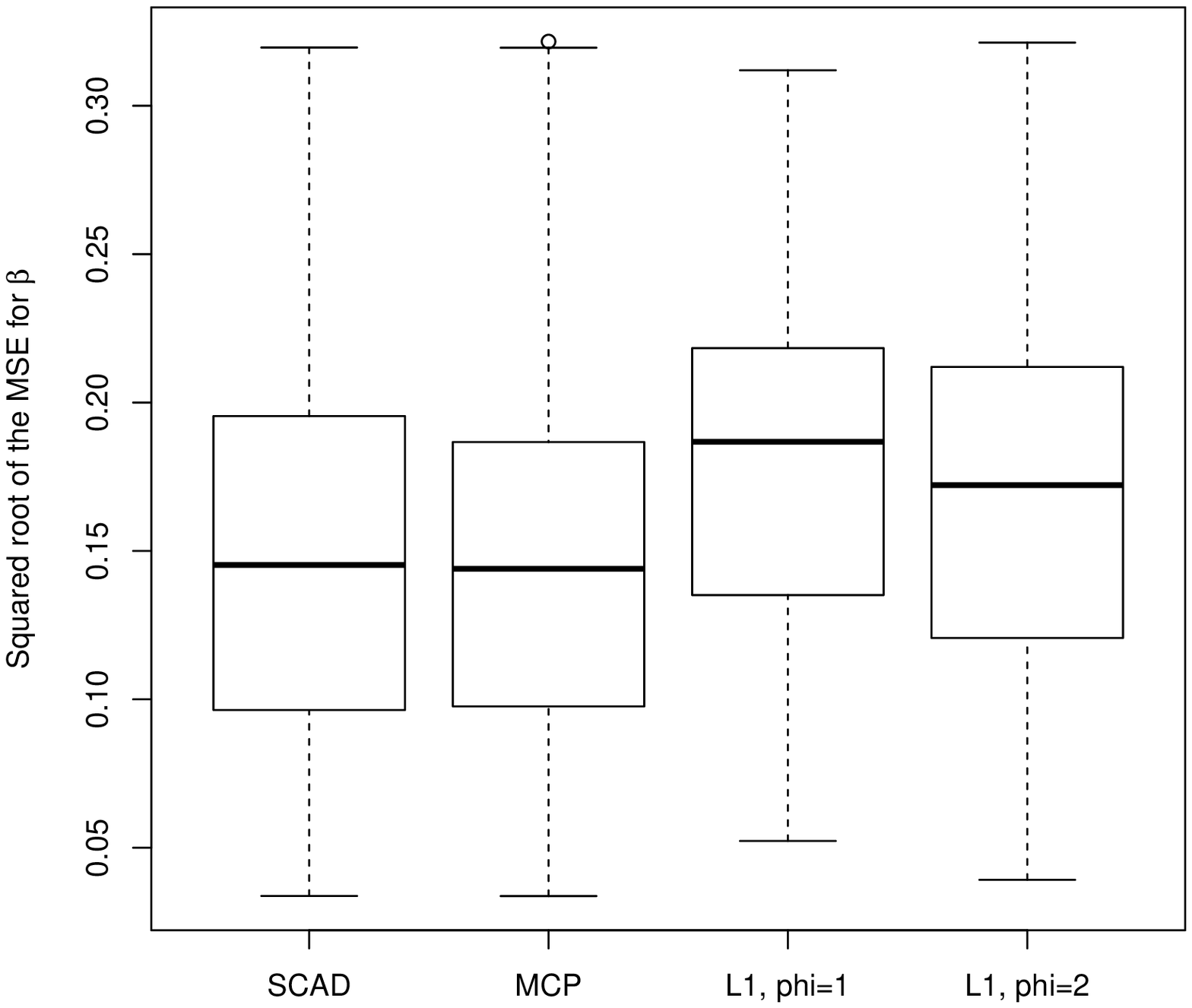} &
\end{array}
$  }
\end{center}
\end{figure}
}

\textbf{Example 3.} We generate data from a homogeneous model given as $%
y_{i}=\mu +\mathbf{x}_{i}^{\text{T}}\mathbf{\boldsymbol{\ \beta }+}\epsilon
_{i},i=1,\ldots ,100$. The predictors, the error term and the coefficients
are simulated in the same way as in Example 1. Let $\mu =2$. We fit the
heterogeneous model (\ref{Mod1}) by using our proposed method. In practice,
we choose the value of $\lambda $ by the modified BIC method as illustrated
in Examples 1 and 2. In this example, for illustration of our penalization
estimation and the subsequent inference method for subgroup identification
in the homogeneous model, we use a set of different values for the tuning
parameter $\lambda$. For small values of $\lambda$, we expect to have more
identified groups. We further conduct inference on heterogeneity between
groups by using the asymptotic normality in Corollary \ref{COR:distribution}.

To test on heterogeneity, we formulate the hypothesis that $\mathcal{H}%
_{0}:\alpha _{1}=(|\widehat{K}|-1)^{-1}\sum_{j=2}^{|\widehat{K}|}\alpha _{j}$
, where $\alpha _{1}$ is the intercept for the largest group, so that we
test on the difference of the intercept for the largest group and the
average intercept for other groups. For demonstration of our inference
procedure applying to the homogeneous model, we choose a set of small values
$\lambda =(0.15,0.20,0.25)$, so that more than one groups are identified by
the penalization procedure. For small values of $\lambda $, since subgroups
with small sizes may be identified, we adjust to estimate $\sigma ^{2}$ by $%
\widehat{\sigma }^{2}=(n-2-p)^{-1}\sum\nolimits_{i=1}^{n}(y_{i}-\widehat{\mu
}_{i}-\mathbf{x}_{i}^{\text{T}}\widehat{\mathbf{\boldsymbol{\beta }}}\mathbf{%
)}^{2}$, where $\widehat{\mu }_{i}=\widehat{\alpha }_{1}$ for $i\in \mathcal{%
G}_{1}$ and $\widehat{\mu }_{i}=(|\widehat{K}|-1)^{-1}\sum_{j=2}^{|\widehat{K%
}|}\widehat{\alpha }_{j}$ otherwise. Figure \ref{FIG:boxplots3} shows the
boxplots of the p-values for the hypothesis testing based on the 100
simulation realizations for different values of $\lambda $. The estimates of
the intercepts are obtained by the MCP and SCAD methods, respectively. We
observe that the median values of the p-values are large in general.

{\normalsize
\begin{figure}[tbp]
\caption{Boxplots of the p-values for the hypothesis testing in Example 3
based on the 100 simulation realizations for different values of $\protect%
\lambda $.}
\label{FIG:boxplots3}{\normalsize \vspace{-0.3cm}  }
\par
\begin{center}
{\normalsize $%
\begin{array}{cc}
\includegraphics[width=3in]{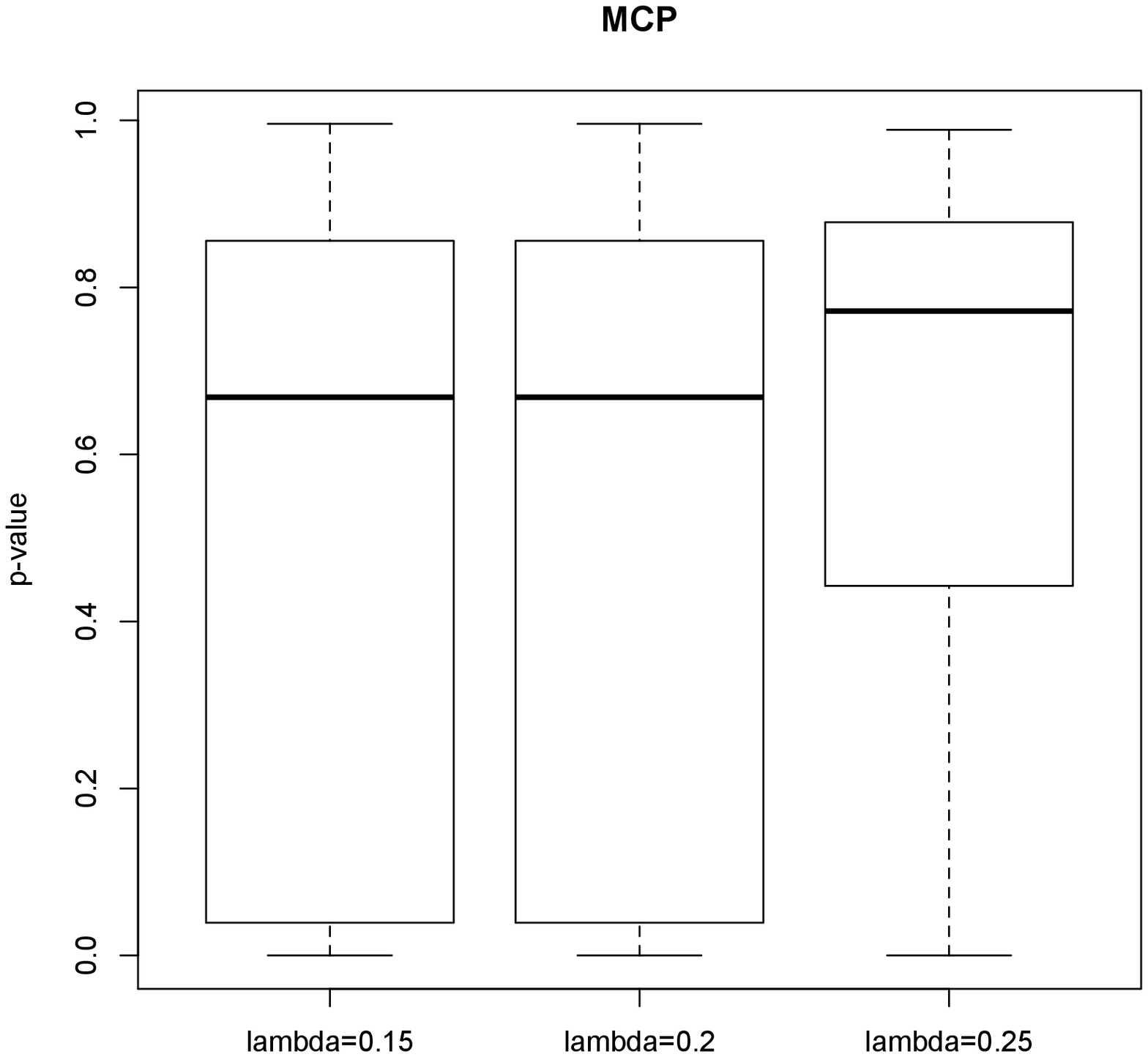} & %
\includegraphics[width=3in]{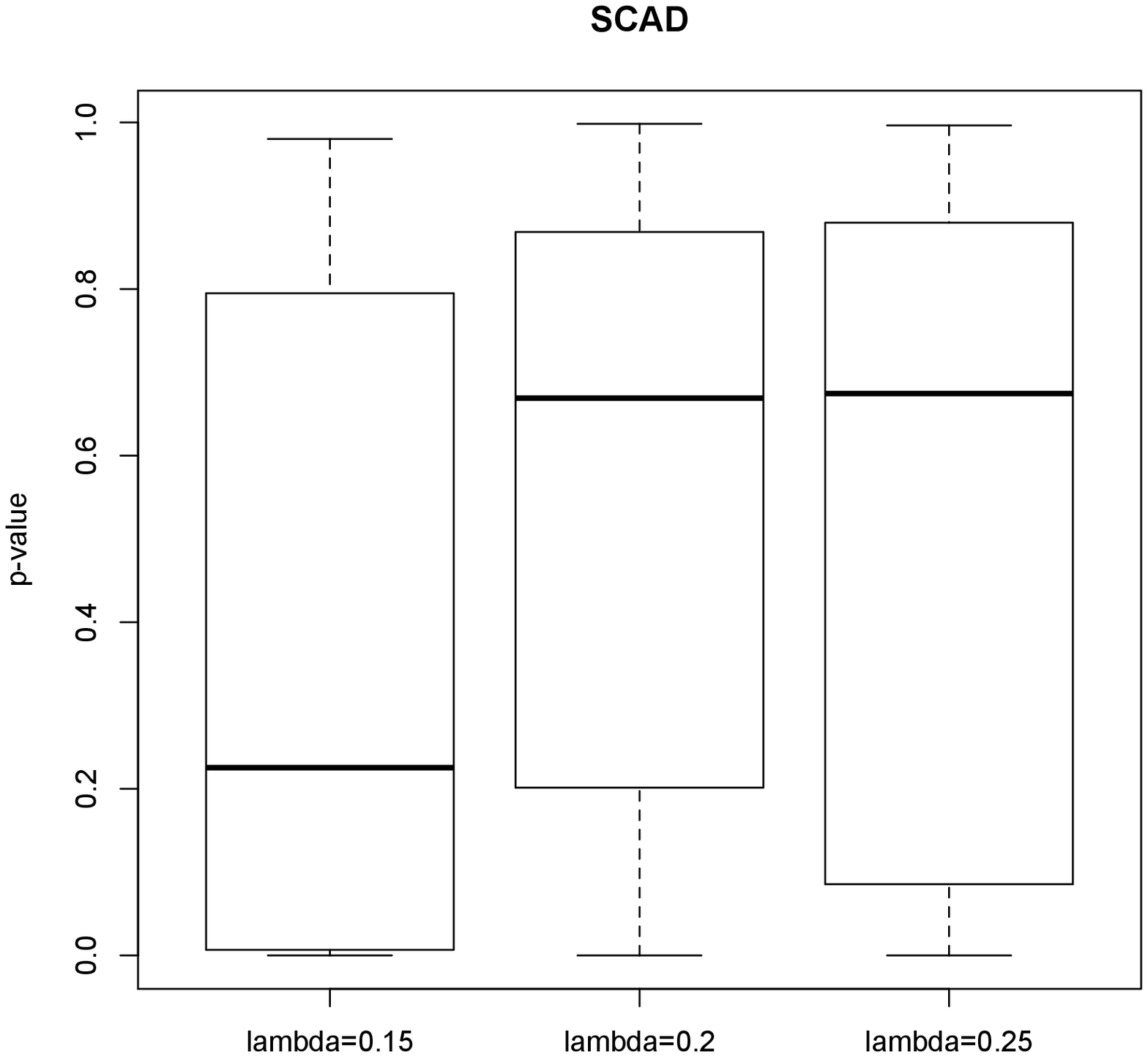}%
\end{array}
$  }
\end{center}
\end{figure}
}

\textbf{Example 4.} \textbf{Case 1. }We simulate data from the same data
generating process as Example 2, so that the data are generated from three
groups with the same size (balanced groups). In this example, we aim to
compare the performance of cluster analysis by using different penalties
including MCP, SCAD, and truncated L$_{1}$ as well as by using the Gaussian
mixture model-based clustering algorithm from the R package of MCLUST (\cite%
{Fraley.Raftery:2002}). In our regression setting, we need to apply MCLUST
to $y_{i}-\mathbf{x}_{i}^{\text{T}}\mathbf{\boldsymbol{\beta }}$ for
cluster analysis. One simple way is to obtain the estimate $\widehat{%
\mathbf{\boldsymbol{\beta }}}$ of $\mathbf{\boldsymbol{\beta }}$ by the
ordinary least squares (OLS) first, and then apply the MCLUST to the pseudo
observations $y_{i}-\mathbf{x}_{i}^{\text{T}}\widehat{\mathbf{\boldsymbol{%
\beta }}}$, which is adopted in our numerical analysis.

For the penalized methods, we apply the same iterative algorithm as
described in Section \ref{SEC:algorithm} to obtain the parameter estimates
by using different penalties. The same BIC method is applied to choose the
tuning parameter as described in Example 2. It is worth noting that the rest steps
remain the same except for the estimation of $\eta _{ij}$, which needs some
modifications due to the use of different penalties. Specifically, for the
truncated L$_{1}$ penalty which has the form $p(|t|,\lambda ;\tau )=\lambda
\min (|t|;\tau )$, where $\tau $ is the thresholding parameter, the estimate
of $\eta _{ij}$ is obtained by minimizing $h(\eta _{ij})=\frac{\vartheta }{2}%
(\delta _{ij}-\eta _{ij}\mathbf{)}^{2}+\lambda \min (|\eta _{ij}|;\tau )$.
We then apply the difference of convex programming technique as given in (%
\cite{Shen.Huang:2010}) to obtain the minimizer of $h(\eta _{ij})$. In this
algorithm, the function $h(\eta _{ij})$ needs to be decomposed into
difference of two convex functions $h_{1}(\eta _{ij})-h_{2}(\eta _{ij})$,
where $h_{1}(\eta _{ij})=\frac{\vartheta }{2}(\delta _{ij}-\eta _{ij}\mathbf{%
)}^{2}+\lambda |\eta _{ij}|$ and $h_{2}(\eta _{ij})=\lambda (|\eta
_{ij}|-\tau )_{+}$. This enables us to approximate $h(\eta _{ij})$ by an
upper convex function at iteration $m+1$ which results in
\begin{equation*}
\widehat{\eta }_{ij}^{(m+1)}=\left\{
\begin{array}{cc}
\widehat{\delta }_{ij}^{(m+1)} & \text{ if }|\widehat{\eta }_{ij}^{(m)}|\geq
\tau  \\
\left( |\widehat{\delta }_{ij}^{(m+1)}|-\lambda /\vartheta \right)
_{+}\left( \widehat{\delta }_{ij}^{(m+1)}/|\widehat{\delta }%
_{ij}^{(m+1)}|\right)  & \text{otherwise}%
\end{array}%
\right. .
\end{equation*}

One important evaluation criterion for clustering methods is based on their
ability to reconstruct the true underlying cluster structure. We, therefore,
use the Rand Index measure (\cite{Rand:1971}) to evalute the accuracy of the
clustering results. The Rand Index is viewed as a measure of the percentage
of correct decisions made by the algorithm. It is computed by using the
formula:%
\begin{equation*}
\text{RI}=\frac{\text{TP}+\text{TN}}{\text{TP}+\text{FP}+\text{FN}+\text{TN}}%
,
\end{equation*}%
where a true positive (TP) decision assigns two observations from the same
ground truth group to the same cluster, a true negative (TN) decision
assigns two observations from different groups to different clusters, a
false positive (FP) decision assigns two observations from different groups
to the same cluster, and a false negative (FN) decision assigns two
observations from the same group to different clusters. \ The Rand Index
lies between 0 and 1. Higher values of the Rand Index indicate better
performance of the algorithm.

Table \ref{TAB:case1EX4} presents the mean and standard error (s.e.) of $%
\widehat{K}$, the square root of the MSE (SMSE) for the estimated $\mathbf{%
\boldsymbol{\mu }}$ and the clustering accuracy (Accuracy) by different
methods. For the truncated L$_{1}$, by taking the same strategy as \cite%
{Shen.Huang:2010}, we use different values $\tau =0.5,1.0,1.5$ for the
thresholding parameter. In the MCLUST\ column, it shows the results by using
the MCLUST package with the number of groups selected by the BIC method
which is the default method in the MCLUST\ package and is widely used for
determining the number of clusters in practice. In the MCLUST-MCP column, it
shows the results by using the MCLUST package with the number of groups
determined by our proposed penalized approach with MCP penalty.{\normalsize
\begin{table}[tbph]
\caption{The mean and standard error (s.e.) of $\protect\widehat{K}$ and the
square root of the MSE (SMSE) for the estimated $\mathbf{\boldsymbol{%
\protect\mu }}$ as well as the clustering accuracy (Accuracy) by different methods
based on 100 realizations with $n=100$ for Case 1 of Example 4 with balanced
groups.}
\label{TAB:case1EX4}
\begin{center}
{\normalsize \hspace{-2cm}
\begin{tabular*}{0.95\textwidth}{@{\extracolsep{\fill}}c|c|c|c|ccc|c|c}
\cline{1-9}
&  & MCP & SCAD & \multicolumn{3}{c|}{Truncated L$_{1}$} & MCLUST &
MCLUST-MCP \\
&  &  &  & $\tau =0.5$ & $\tau =1.0$ & $\tau =1.5$ &  &  \\ \cline{1-9}
$K$ & mean & 3.570 & 3.600 & 6.930 & 3.960 & 2.390 & 2.400 & --- \\
& s.e. & 0.671 & 0.696 & 0.956 & 0.887 & 0.737 & 0.711 & --- \\
SMSE of $\mu $ & mean & 0.589 & 0.585 & 0.597 & 0.605 & 0.963 & 0.791 & 0.607
\\
& s.e. & 0.157 & 0.154 & 0.158 & 0.164 & 0.195 & 0.380 & 0.134 \\
Accuracy & mean & 0.897 & 0.892 & 0.829 & 0.873 & 0.707 & 0.777 & 0.864 \\
& s.e. & 0.059 & 0.057 & 0.066 & 0.064 & 0.112 & 0.193 & 0.058 \\ \cline{1-9}
\end{tabular*}
}
\end{center}
\end{table}
}

From Table \ref{TAB:case1EX4}, we observe that the proposed concave fusion
penalized methods, MCP and SCAD, have better performance than other methods.
They have higher clustering accuracy rates and smaller SMSE values for $%
\widehat{\mathbf{\boldsymbol{\mu }}}$ than others. This result is further
reflected by the boxplots in Figure \ref{FIG:boxplotCase1EX4} of accuracy
rates for the MCP, SCAD, truncated L$_{1}$ with $\tau =1.0$, and MCLUST
methods. For the truncated L$_{1}$, it has the best performance at $\tau =1.0
$ among the three different values for $\tau $. Moreover, the three
penalized methods, MCP, SCAD and truncated L$_{1}$ with $\tau =1.0$, can
identify the cluster membership more correctly than the MCLUST method by
observing higher accuracy rates. The MCP\ improves the accuracy rate by $%
15.4\%$ compared to the MCLUST. It is worth noting that in order to apply
the Gaussian mixture model-based method, how many clusters to be used is
always crucial. For the MCLUST-MCP, instead of using the BIC, we use our
proposed penalized MCP approach to determine the number of clusters and then
apply the MCLUST, we see that the accuracy rate is improved compared to the
MCLUST with the BIC\ method. This result indicates that our proposed concave
penalized method also provides a possible tool to determine the number of
clusters for the Gaussian mixture model-based method.{\normalsize
\begin{figure}[tbp]
\caption{Boxplots of the clustering accuracy for the MCP, SCAD, truncated $%
L_{1}$ and MCLUST based on the 100 simulation realizations in Case 1 of
Example 4.}
\label{FIG:boxplotCase1EX4}{\normalsize \vspace{-0.3cm}  }
\par
\begin{center}
{\normalsize $\includegraphics[width=3in]{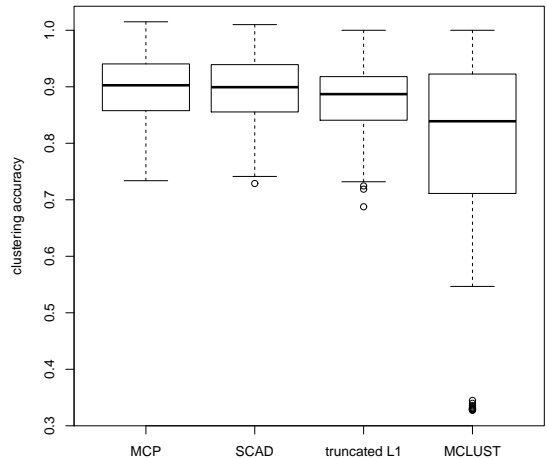} $  }
\end{center}
\end{figure}
}

\textbf{Case 2. }In this setting, we generate data from three groups with
different sizes (unbalanced groups). We consider two simulation designs:
Design 1: $\mu _{i}$'s are generated from three different values $-2$, $0$, $%
2$ with probabilities 0.2, 0.3, 0.5, respectively, and Design 2: $\mu _{i}$%
's are generated from $-2$, $0$, $2$ with probabilities 0.1, 0.3, 0.6,
respectively. Other terms are simulated according to the same setting as
Case 1. Table \ref{TAB:case2EX4} presents the mean and standard error
(s.e.) of $\widehat{K}$, the square root of the MSE (SMSE) for the estimated
$\mathbf{\boldsymbol{\mu }}$ and the clustering accuracy (Accuracy) by the
MCP, SCAD, truncated L$_{1}$, MCLUST and MCLUST-MCP\ based on 100
realizations. We see that for the MCP, SCAD and truncated L$_{1}$ methods,
the performance for the two unbalanced designs is comparable to that for the
balanced design in Case 1. Again the MCP\ and SCAD outperform the other
methods. The performance of MCLUST-MCP shows improvement over MCLUST. \ For
the MCLUST, the estimated number of groups $\widehat{K}$ decreases as the
design becomes more unbalanced. The smallest group is not successfully
identified for most replications. For the penalized method, however, the $%
\widehat{K}$ values remain similar for different designs. Hence, the MCLUST\
may be more sensitive to cluster sizes based on these simulation results.%
{\normalsize
\begin{table}[t]
\caption{The mean and standard error (s.e.) of $\protect\widehat{K}$ and the
square root of the MSE (SMSE) for the estimated $\mathbf{\boldsymbol{%
\protect\mu }}$ as well as the clustering accuracy (Accuracy) by different methods
based on 100 realizations with $n=100$ for Case 2 of Example 4 with
unbalanced groups.}
\label{TAB:case2EX4}
\begin{center}
{\normalsize \hspace{-2cm}
\begin{tabular*}{0.95\textwidth}{@{\extracolsep{\fill}}c|c|c|c|ccc|c|c}
\cline{1-9}
&  & MCP & SCAD & \multicolumn{3}{c|}{Truncated L$_{1}$} & MCLUST &
MCLUST-MCP \\
&  &  &  & $\tau =0.5$ & $\tau =1.0$ & $\tau =1.5$ &  &  \\ \cline{1-9}
\multicolumn{9}{c}{Design 1} \\ \cline{1-9}
$K$ & mean & 3.730 & 3.660 & 6.540 & 3.870 & 2.360 & 2.380 & --- \\
& s.e. & 0.670 & 0.713 & 1.049 & 0.928 & 0.659 & 0.663 & --- \\
SMSE of $\mu $ & mean & 0.561 & 0.556 & 0.585 & 0.577 & 0.893 & 0.771 & 0.592
\\
& s.e. & 0.126 & 0.130 & 0.127 & 0.146 & 0.135 & 0.309 & 0.124 \\
Accuracy & mean & 0.890 & 0.891 & 0.822 & 0.872 & 0.733 & 0.792 & 0.846 \\
& s.e. & 0.048 & 0.048 & 0.051 & 0.058 & 0.092 & 0.152 & 0.064 \\ \cline{1-9}
\multicolumn{9}{c}{Design 2} \\ \cline{1-9}
$K$ & mean & 3.700 & 3.730 & 6.350 & 3.960 & 2.690 & 2.230 & --- \\
& s.e. & 0.717 & 0.709 & 0.880 & 1.197 & 1.473 & 0.679 & --- \\
SMSE of $\mu $ & mean & 0.488 & 0.487 & 0.522 & 0.502 & 0.852 & 0.763 & 0.579
\\
& s.e. & 0.121 & 0.120 & 0.122 & 0.127 & 0.135 & 0.220 & 0.148 \\
Accuracy & mean & 0.898 & 0.899 & 0.823 & 0.877 & 0.713 & 0.793 & 0.818 \\
& s.e. & 0.048 & 0.047 & 0.056 & 0.054 & 0.118 & 0.123 & 0.097 \\ \cline{1-9}
\end{tabular*}
}
\end{center}
\end{table}
}

\section{Empirical example\label{SEC:applications}}

In this section, we use the Cleveland Heart Disease Dataset to illustrate
our method. This dataset is available at the UCI machine learning
repository. The dataset has 13 clinical measurements on 297 individuals. As
described in \cite{Lauer:1999}, the maximum heart rate achieved (thalach)
variable is related to cardiac mortality. In addition, some categorical
variables are also used to check heart problems including chest pain type,
exercise induced angina indicator, ST depression induced by exercise
relative to rest, slope of the peak exercise ST segment, number of major
vessels colored by fluoroscopy and the heart status (normal=3; fixed
defect=6; reversible defect=7). We use the fitted value of thalach as the
response variable by projecting it onto the linear space spanned by the
categorical variables. Our interest is to conduct subgroup analysis for the fitted value of thalach as the response $y$ after adjusting for the effects of the covariates: $x_{1}=$age in years;
$x_{2}=$gender; $x_{3}=$resting blood pressure; $x_{4}=$serum cholesterol; $%
x_{5}=$fasting blood sugar indicator; and $x_{6}=$resting
electrocardiographic results.

We first plot the kernel density estimates of $y_{i}-\mathbf{x}_{i}^{\text{ T%
}}\widehat{\boldsymbol{\beta }}$ in Figure \ref{FIG:density}, where $%
\widehat{\boldsymbol{\beta }}$ is obtained from OLS estimation. Clearly, we
see that after adjusting for the effects of the covariates, the distribution
in Figure \ref{FIG:density} still shows multiple modes. The heterogeneity
may be caused by some unobserved latent factors. Hence, it is not suitable
to fit a standard linear regression model with a common intercept by using
the response and the predictors. Instead we fit the heterogeneous model $%
y_{i}=\mu _{i}+\mathbf{x}_{i}^{\text{ T}}\boldsymbol{\beta }\mathbf{+}%
\epsilon _{i},i=1,\ldots ,297$, and we identify subgroups by our proposed ADMM algorithm. We select the tuning parameter by minimizing the
modified BIC in a certain range by following the same rule as given in
Example 2 of Section \ref{SEC:examples}. As a result, two major groups are
identified by both of the MCP\ and SCAD methods. We also conduct inference
by testing the difference of the intercepts for the two identified groups by
using the asymptotic normality in Corollary \ref{COR:distribution}, and we
find that the p-values are close to zero for both of the MCP and SCAD
methods.

{\normalsize
\begin{figure}[tbp]
\caption{Density plot of the response variable after adjusting for the
effects of the covariates for the empirical example.}
\label{FIG:density}{\normalsize \vspace{-0.3cm}  }
\par
\begin{center}
{\normalsize $\includegraphics[width=3in]{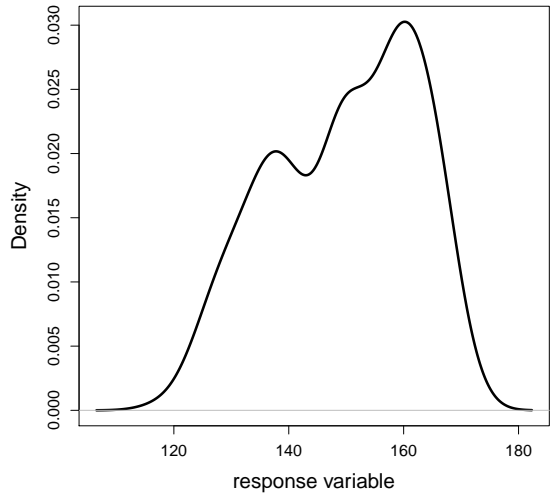} $  }
\end{center}
\end{figure}
} {\normalsize
\begin{table}[tbph]
\caption{The estimated values (est) for the coefficients $\mathbf{%
\boldsymbol{\protect\beta }}$, their standard deviations (s.d.) and the
p-values for testing the significance of the coefficients by the OLS, MCP
and SCAD, respectively.}
\label{TAB:coefficient}
\begin{center}
{\normalsize
\begin{tabular*}{0.95\textwidth}{l|l|@{\extracolsep{\fill}}cccccc}
\hline
&  & $\boldsymbol{\beta} _{1}$ & $\boldsymbol{\beta} _{2}$ & $\boldsymbol{%
\beta} _{3}$ & $\boldsymbol{\beta} _{4}$ & $\boldsymbol{\beta} _{5}$ & $%
\boldsymbol{\beta} _{6}$ \\ \hline
& est & $-0.345$ & $-4.120$ & $-0.028$ & $-0.008$ & $0.183$ & $-1.359$ \\
OLS & s.d. & $0.083$ & $1.534$ & $0.042$ & $0.0142$ & $2.031$ & $0.725$ \\
& p-value & $<0.001$ & $0.007$ & $0.502$ & $0.563$ & $0.928$ & $0.061$ \\
\hline
& est & $-0.355$ & $-3.825$ & $-0.007$ & $-0.006$ & $0.628$ & $-1.849$ \\
MCP & s.d. & $0.040$ & $0.752$ & $0.021$ & $0.007$ & $1.016$ & $0.354$ \\
& p-value & $<0.001$ & $<0.001$ & $0.563$ & $0.558$ & $0.283$ & $<0.001$ \\
\hline
& est & $-0.358$ & $-3.698$ & $-0.012$ & $-0.004$ & $1.091$ & $-2.129$ \\
SCAD & s.d. & $0.040$ & $0.743$ & $0.021$ & $0.007$ & $1.005$ & $0.351$ \\
& p-value & $<0.001$ & $<0.001$ & $0.558$ & $0.554$ & $0.278$ & $<0.001$ \\
\hline
\end{tabular*}
}
\end{center}
\end{table}
} In Table \ref{TAB:coefficient} we report the estimated coefficients $%
\widehat{\mathbf{\boldsymbol{\beta }}}$, their standard deviations (s.d.)
and the p-values for testing the significance of the coefficients by the
proposed method with the MCP and SCAD pairwise fusion, and the OLS
estimation by assuming a common intercept. The standard deviation for the
MCP and SCAD methods is calculated by the asymptotic formula given in
Corollary \ref{COR:distribution}. The age and gender variables show a
strongly significant effect by these three methods with p-values close to
zero, while resting blood pressure and serum cholesterol show a very weak
effect by the three methods with large p-values. Moreover, by the MCP and
SCAD methods, the effects of fasting blood sugar indicator and resting
electrocardiographic results become more significant than the results by the
OLS method. This result indicates that by recovering the hidden
heterogeneous structure of the data, it helps us identify useful variables
which may have effects on the response. We also calculate the coefficient of
determination $R^{2}$, and obtain $R^{2}=0.667$, $0.704$ and $0.109$ for
MCP, SCAD and OLS methods. We see that taking into account the subgroup
structure 
leads to a significant improvement of the model fitting. Next we apply the
Gaussian mixture model-based method to this dataset for cluster analysis. As
described in Example 4 of the simulation section, we apply the MCLUST\ to
the pseudo observations $y_{i}-\mathbf{x}_{i}^{\text{T}}\widehat{\mathbf{%
\boldsymbol{\beta }}}$, where $\widehat{\mathbf{\boldsymbol{\beta }}}$ is
obtained from the OLS. As a result, two subgroups are identified. For the
real data, since the true underlying cluster structure is unknown, we cannot
use the external criterion, Rand Index measure, to evaluate and compare
different methods. Instead, we use the internal criterion, the
Davies--Bouldin index, to assess the quality of clustering algorithms, which
is calculated by the formula: DB$=\widehat{K}^{-1}\sum $ $_{k=1}^{\widehat{K}%
}\max_{k^{\prime }\neq k}((\sigma _{k}+\sigma _{k^{\prime
}})/d(c_{k},c_{k^{\prime }}))$, where $\widehat{K}$ is the estimated number
of clusters, $c_{x}$ is the centroid of cluster $x$, $\sigma _{x}$ is the
average distance of all observations $y_{i}-\mathbf{x}_{i}^{\text{T}}%
\widehat{\mathbf{\boldsymbol{\beta }}}$ in cluster $x$ to centroid $c_{x}$,
and $d(c_{k},c_{k^{\prime }})$ is the distance between centroids $c_{k}$ and
$c_{k^{\prime }}$. The clustering algorithm that has the smallest
Davies--Bouldin index is considered the best algorithm based on this
criterion. The Davies--Bouldin index values for MCP, SCAD and MCLUST\ are
0.469, 0.467, and 0.506, respectively, so that the MCP and SCAD outperform
the MCLUST\ based on this criterion.

\section{Discussion\label{SEC:Discussion}}


The model (\ref{Mod1}) is related to the Neyman-Scott models (\cite%
{NeymanScott:1948}). In the terminology of Neyman and Scott, the $\mu_i$'s
in (\ref{Mod1}) are called incident parameters. In the literature, such
parameters are usually treated as nuisance parameters, while the main
interest lies in estimating the common parameter such as $\{\boldsymbol{\beta}%
, \sigma^2\}$ in (\ref{Mod1}) based on panel data (\cite{Lancaster:2000}).
The problem we consider here is different and we use the $\mu_i$'s to
represent latent heterogeneity in the observations for the purpose of conducting
subgroup analysis. Also we do not assume that panel data are
available, so model (\ref{Mod1}) is not identifiable without a constraint on
the parameter space such as the subgroup structure considered in the
present paper.

It is also possible to adopt a random effects model approach by taking the $%
\mu _{i}$'s in (\ref{Mod1}) as random variables from a mixture distribution.
Then the estimation and inference can be carried out using a
likelihood-based method. The main difficulty in applying this approach is
that it requires specifying the number of subgroups, the parametric form of
the mixture distribution, and an assumption on the
error distribution. It is worth noting that the choice of the number of groups is
always crucial in mixture model-based methods. Different methods on
this topic have been proposed in the literature. Among them, the
Bayesian model selection criteria (\cite{Fraley.howmany:1998}) are widely
used, and the gap statistic proposed in \cite{Tibshirani.et:2001} is also an
important tool. Our proposed penalized method provides another possible
approach to automatically estimate the number of groups with reliable theoretical properties. By using the MCLUST, our simulation studies show
that the clustering accuracy is improved by using the proposed penalized
method to select the number of groups compared to the BIC.

In our theoretical results, we allow $p$, the dimension of the regression
parameter $\boldsymbol{\beta}$, to diverge with $n$, but require it to be
smaller than $n$. For models and data with $p > n$, a sparsity condition
needs to be imposed on $\boldsymbol{\beta}$ and an additional penalty term
to enforce the sparsity is required. Computationally, we can still derive an
algorithm within the framework. However, much extra effort is needed to
establish the theoretical properties of the estimators in this
high-dimensional setting. This is an interesting and challenging technical
problem and deserves further investigation, but is beyond the scope of this paper.

The proposed method can be extended to other models including the
generalized linear models and regression models for censored survival data.
Although these extensions appear to be conceptually straightforward, it is a
nontrivial task to develop computational algorithms and establish
theoretical properties in these more complicated models.

\end{document}